\documentclass[journal]{IEEEtran}
\usepackage{amsfonts}
\usepackage{amssymb}
\usepackage{cite}
\usepackage[cmex10]{amsmath}
\usepackage{float}
\usepackage{color}
\usepackage{stfloats,fancyhdr}
\usepackage{amsmath}
\usepackage{amsthm}
\usepackage{enumerate}
\usepackage{array}
\usepackage{algorithm}
\usepackage{algorithmic}
\usepackage{multirow}
\usepackage{changepage}
\usepackage[normalem]{ulem}

\usepackage{url}
\usepackage{xr-hyper}
\usepackage{hyperref}

\makeatletter
\newcommand*{\addFileDependency}[1]{
  \typeout{(#1)}
  \@addtofilelist{#1}
  \IfFileExists{#1}{}{\typeout{No file #1.}}
}
\makeatother

\newtheorem{remark}{Remark}

\IEEEoverridecommandlockouts

\ifCLASSINFOpdf
  \usepackage[pdftex]{graphicx}
  \DeclareGraphicsExtensions{.pdf,.jpeg,.png}
\else
  \usepackage[dvips]{graphicx}
  \DeclareGraphicsExtensions{.pdf}
\fi

\usepackage{subfigure}

\usepackage{fancybox,dashbox}

\UseRawInputEncoding

\begin{document}
%
\title{\huge{Design and Implementation of Time-Sensitive Wireless IoT Networks on Software-Defined Radio}}

\author{Jiaxin~Liang,~\IEEEmembership{Student Member,~IEEE,}
        He~Chen,~\IEEEmembership{Member,~IEEE,}
        and~Soung~Chang~Liew,~\IEEEmembership{Fellow,~IEEE}%
\thanks{J. Liang, H. Chen, and S. C. Liew are with Department of Information Engineering, The Chinese University of Hong Kong, Hong Kong SAR, China (email: \{lj015, he.chen, soung\}@ie.cuhk.edu.hk).}}

\maketitle

\begin{abstract}
Time-sensitive wireless networks are an important enabling building block for many emerging industrial Internet of Things (IoT) applications. Quick prototyping and evaluation of time-sensitive wireless technologies are desirable for R\&D efforts. Software defined radio (SDR), by allowing wireless signal processing on a personal computer (PC), has been widely used for such quick prototyping efforts. Unfortunately, because of the \textit{uncontrollable delay} between the PC and the radio board, SDR is generally deemed not suitable for time-sensitive wireless applications that demand communication with low and deterministic latency. For a rigorous evaluation of its suitability for industrial IoT applications, this paper conducts a quantitative investigation of the synchronization accuracy and end-to-end latency achievable by an SDR wireless system. To this end, we designed and implemented a time-slotted wireless system on the Universal Software Radio Peripheral (USRP) SDR platform. We developed a time synchronization mechanism to maintain synchrony among nodes in the system. To reduce the delays and delay jitters between the USRP board and its PC, we devised a {\textit{Just-in-time}} algorithm to ensure that packets sent by the PC to the USRP can reach the USRP just before the time slots they are to be transmitted. Our experiments demonstrate that $90\%$ ($100\%$) of the time slots of different nodes can be synchronized and aligned to within $ \pm 0.5$ samples or $ \pm 0.05\mu s$ ($ \pm 1.5$ samples or $ \pm 0.15\mu s$), and that the end-to-end packet delivery latency can be down to $3.75ms$. This means that SDR-based solutions can be applied in a range of IIoT applications that require tight synchrony and moderately low latency, e.g., sensor data collection, automated guided vehicle (AGV) control, and Human-Machine-Interaction (HMI).
\end{abstract}



\IEEEpeerreviewmaketitle



\section{Introduction}
\textcolor{black}{The Industrial Internet of Things (IIoT) apply IoT technologies in the industrial domain. IIoT has attracted great attention from governments, academia, and industry, thanks to its potential to boost efficiency and enhance flexibility in future smart factories \cite{yLiao_iiot_2018}. The industry giant GE pointed out that providing powerful and pervasive connectivity between machines, workers and materials in factories will be essential to unlocking the full potential of IIoT \cite{everything_iiot}. }

Connectivity between devices in an industrial environment has until now been dominated by wired communication. Replacing the wired communication infrastructure in today's factories by its wireless counterpart will bring many benefits, including reduced installation and maintenance costs, quick reconfiguration, and mobility \cite{Huang2018, Hellstrom_SDR}. However, simple installation of current wireless technologies such as WiFi and 4G, in the industrial environment will not yield satisfactory performance \cite{luvisotto2016ultra}. Typical industrial applications require deterministic real-time exchange of small amounts of data (e.g., a single control command) with tight latency constraints, whereas modern wireless communication systems have been engineered for the exchange of large amounts of data with loose requirements on synchrony and timeliness \cite{chen2018}. To close this gap, conventional solutions that rely on general-purpose wireless chipsets may need to be replaced with dedicated solutions \cite{Hellstrom_SDR} with customized wireless physical and data-link layer designs tailored for time-sensitive industrial applications.

Software-defined radio (SDR), widely studied in the past few decades, is an appealing alternative to conventional radio for R\&D efforts \cite{Hellstrom_SDR,jondral2005software}. The main goal of SDR is to facilitate the implementation of radio signal processing components, traditionally done on customized hardware (e.g., equalizers, modulators, and coders), by software on general-purpose computers such as PCs. The softwarization can significantly shorten the development and evaluation cycle of new radio techniques. 

Existing development efforts on SDR platforms have been focusing on consumer wireless technologies (e.g., IEEE 802.15.4 \cite{bloessl2013gnu}, IEEE 802.11a \cite{tan2011sora,drozdenko2017hardware}, IEEE 802.11ac  MIMO \cite{wu2017tick}, and standard encoder/decoder modules \cite{torrego2012data}). Due to the indeterminate delays in the SDR architecture (e.g., processing delay in the PC, and transmission delay between the PC and the radio board, are random), the ``PC + radio board'' SDR structure has been widely deemed not suitable for time-sensitive IIoT applications. However, there has been no quantitative evaluation to quantify the extent to which this impression is true – i.e., there are no systematic study on the range of deterministic delays achievable by the ``PC + radio board'' SDR architecture. \textcolor{black}{There have been several recent IIoT investigations \cite{tongyang_2020_iot, tschimben_ieee_2019, mathur2017demo} that leverage PC-based SDR to prove concepts and to demonstrate system capabilities.}

For time-critical applications, a new model-based SDR approach has been introduced, where software tools are used to automatically translate high-level models to low-level hardware description language (HDL) \cite{Hellstrom_SDR}. These tools enable the replacement of the general-purpose PC by the Field-programmable gate array (FPGA) embedded systems for fast signal processing. Though the original goal of the model-based SDR is to allow designers to focus on system-level designs only, efficient implementation and debugging of the designs still demand a deep understanding of FPGA and its programming. Model-based SDR is much harder to handle than PC-based SDR, especially for people with software programming background only.

Meanwhile, a new trend, initiated by the O-RAN Alliance made up of major worldwide mobile operators and computing platform manufactures, is to develop commercial radio-access network (RAN) products based on a ``PC + radio board'' architecture similar to PC-based SDR \cite{noauthor_membership_nodate, bonati_open_2020}. For real-time performance, the O-RAN Alliance promotes the use of powerful computing platforms and real-time operating systems (OS).

This new trend motivates us to revisit an unanswered question: How to efficiently enhance the performance of the PC-based SDR, in terms of its synchronization precision and end-to-end latency, to allow it to support time-sensitive IIoT applications? To answer this question, we designed and implemented a real-time time-slotted wireless system on the USRP SDR radio platform connected to general-purpose PCs running non-real-time OS. We refer to our system as RTTS-SDR (Real-Time Time-slotted System over SDR). 

There are two fundamental challenges to the design of RTTS-SDR. The first challenge is the time synchronization challenge: to align the time slot boundaries of different nodes without the nodes being physically connected to a common external clock source, subject to the random delays between the USRP and the PC of the nodes. The second challenge is the low-latency challenge:  to ensure the USRP of a node will transmit a packet in a near-future time slot specified by its PC, again subject to the random delay between the PC (where the signal samples and instructions are generated) and the USRP (where the signal samples are transmitted).

RTTS-SDR has several salient features: (1) It incorporates a time-synchronization mechanism to maintain microsecond-level synchrony among nodes. (2) It uses a {\textit{Just-in-time}} algorithm to ensure that, at the transmitter side, the PC generates and sends a packet to the USRP just a little ahead of the transmission time of the packet at the USRP, thereby reducing the end-to-end delivery latency. (3) It is a complete TCP/IP compatible system ready to run any TCP/IP application. (4) It can be reconfigured for different latency-throughput requirements by changing PHY and data-link layer parameters.

\textcolor{black}{We believe that RTTS-SDR is a valuable tool upon which other researchers can quickly build and experiment with advanced wireless communications systems, thanks to the flexibility and short development cycle of PC-based SDR. For example, multiuser systems such as OFDMA \cite{yin_ofdma_2006} and Physical-layer Network Coding (PNC) \cite{zhang_hot_2006} in which multiple users transmit in the same and carefully aligned time slot can use the facility already provided by RTTS-SDR.}

Experiments on RTTS-SDR demonstrate that $90\%$ ($100\%$) of the slot boundaries of different nodes can be synchronized to within $ \pm 0.5$ samples or $ \pm 0.05\mu s$ ($ \pm 1.5$ samples or $ \pm 0.15\mu s$), and that packets of $36$ bytes can be delivered with deterministic end-to-end latency of $3.75ms$. RTTS-SDR currently runs on a generic Linux OS---the latency can be further reduced if it is deployed on a real-time OS.

\section{Related Work}
\textcolor{black}{The authors of WirelessHP \cite{luvisotto2017physical} implemented a customized PHY layer on USRP with optimized short OFDM packets for industrial usage. The implementation makes use of offline rather than real-time signal processing and is therefore not ready to run real applications. Also, \cite{luvisotto2017physical} did not address the MAC design to coordinate channel access by nodes. RTTS-SDR, by contrast, is a real-time system that supports real applications, with precise synchronization among nodes. Furthermore, rather than a one-size-fits-all system, RTTS-SDR can be reconfigured with different OFDM parameters for different applications.}

\textcolor{black}{The authors of  \cite{nikumani_LSB_2019} proposed a beacon-based synchronization method to synchronize the nodes in an IEEE 802.15.4 cluster-tree network. However, \cite{nikumani_LSB_2019} focused on the scalability and the overhead of the algorithm, with the synchronization precision largely overlooked. Also, the platform does not support real-time applications yet.}

\textcolor{black}{Recently, \cite{seijo_w-sharp_2020} proposed a customized wireless system, w-SHARP, to enhance the 802.11 PHY and MAC layers. Specifically, \cite{seijo_w-sharp_2020} attempted to reduce cycle time by optimizing the overhead (e.g. InterFrame Spacing) in both the PHY and MAC layers. An ARM+FPGA SoC-based prototype was built to demonstrate the concept. RTTS-SDR, by contrast, is implemented	on the ``PC+USRP'' platform. We show that despite the variable delay between PC and USRP, precise synchronization and moderately low end-to-end latency between nodes can be achieved.}


\section{System Architecture}
\begin{figure}
\centering \scalebox{0.35}{\includegraphics{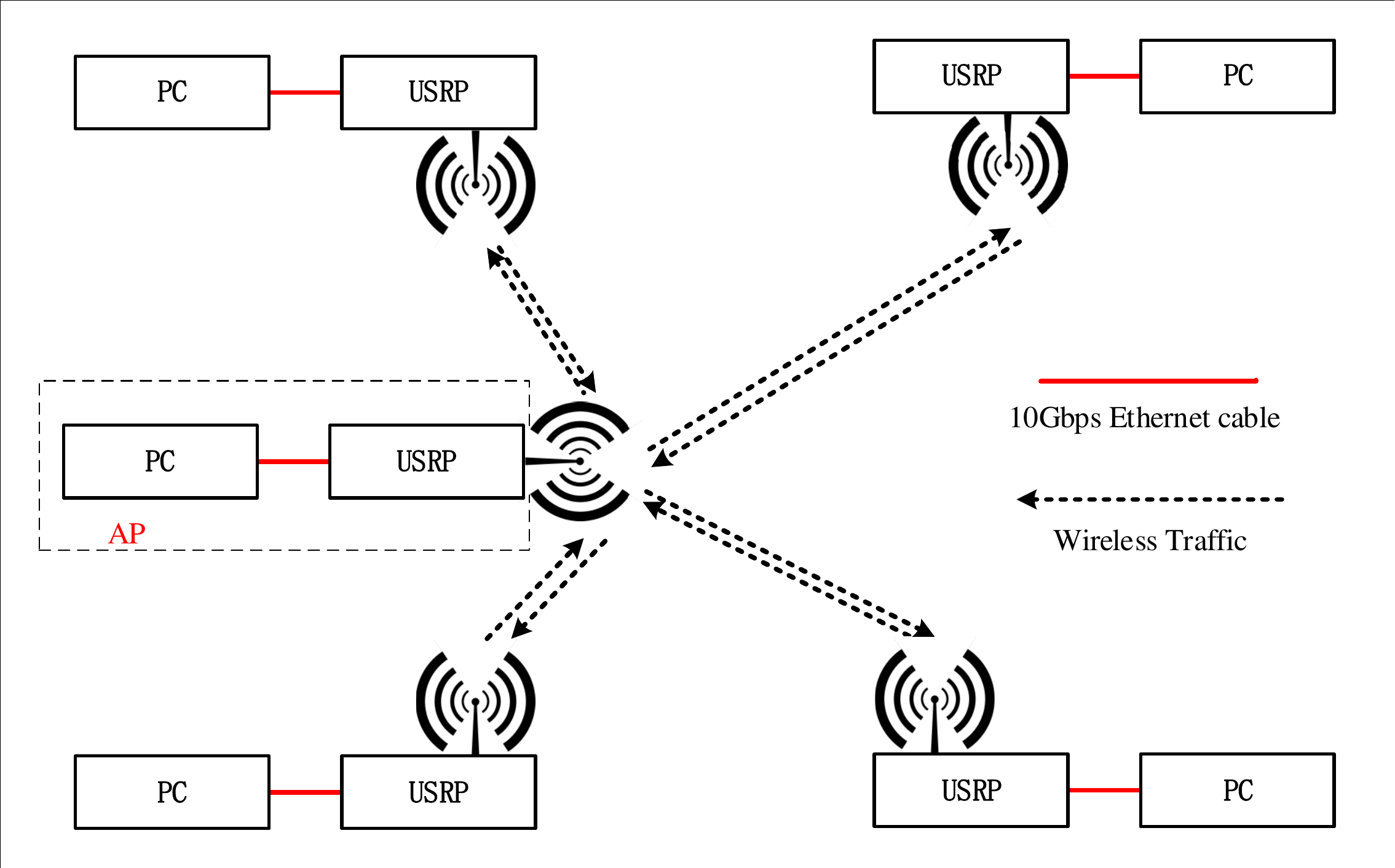}}
\caption{An example of a 5-node system. One of the nodes serves as the AP and other nodes serve as the IoT devices. The AP is responsible for synchronization.}\label{fig:system_arch}
\end{figure}

\subsection{System architecture}
Fig. \ref{fig:system_arch} shows an example of our system with $5$ nodes. Each node consists of a PC and a USRP interconnected by a $10$Gbps Ethernet cable. One of the nodes is the AP, which is also the synchronization coordinator of the overall system. The other nodes are IoT devices. Specifically, the AP's USRP clock gives the system's global reference time. We use $i$ to represent the index of a node. The AP has index $0$ (i.e. $i = 0$). 

\begin{figure}
\centering \scalebox{0.39}{\includegraphics{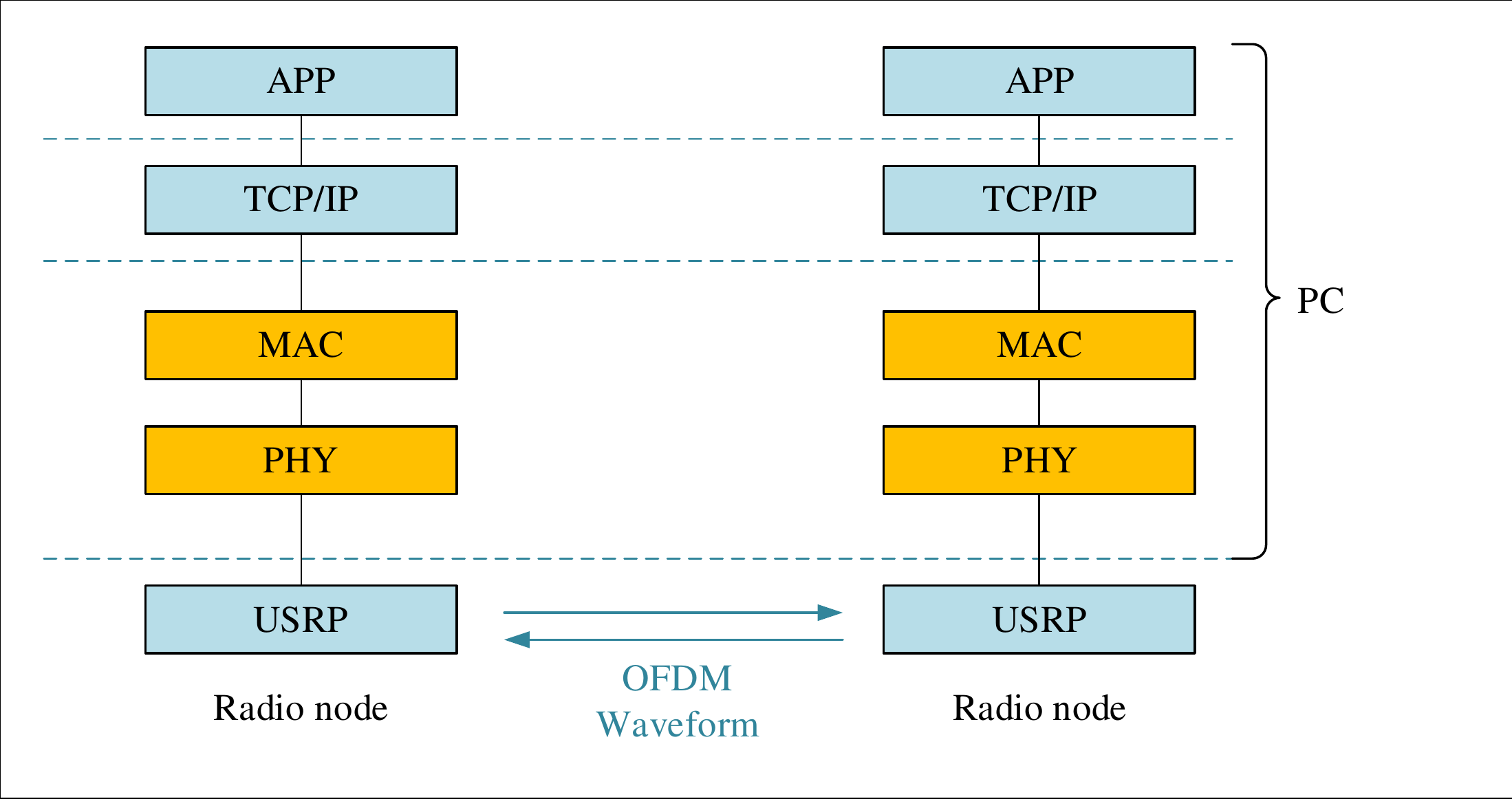}}
\caption{RTTS-SDR provides the complete design for the orange blocks, i.e., MAC, and PHY.}\label{fig:OSIModel}
\end{figure}

RTTS-SDR provides the functionalities of the Physical (PHY) layer, the Data link layer, and the Network layer. Fig. \ref{fig:OSIModel} shows the communication through the layers between two nodes. Users run applications (APP layer) over the TCP/IP layer. The TCP/IP layer then puts the data into or extract the data from the MAC layer and the layers below. RTTS-SDR provides the orange blocks (MAC, and PHY) that are fully compatible with the existing TCP/IP layer in the OS. 

To ensure TCP/IP compatibility, we use the TAP device \cite{TUNTAP_doc} in the OS as a virtual network card to interact with the TCP/IP layer. For the TX path, when an IP datagram is forwarded to the virtual network card, the TAP device generates an Ethernet frame with the IP datagram as the payload and forwards the Ethernet frame to the program that created the TAP device.  GNURadio is the program that creates the TAP device in our system. GNURadio then generates the baseband samples based on the content of the Ethernet frame and sends them to the USRP. For the RX path, GNURadio processes the baseband samples coming from the USRP and then forwards the processed data to the TAP device.

The MAC layer in our system adopts a time-slotted medium access scheme. Due to the random delay jitters between the PC and the USRP, timing control is challenging in the time-slotted system. In particular, when the PC instructs the USRP to transmit a particular frame in a certain time slot, the PC must ensure this frame can arrive at the USRP and be processed at the USRP before the time slot.

\begin{figure}
\centering \scalebox{0.35}{\includegraphics{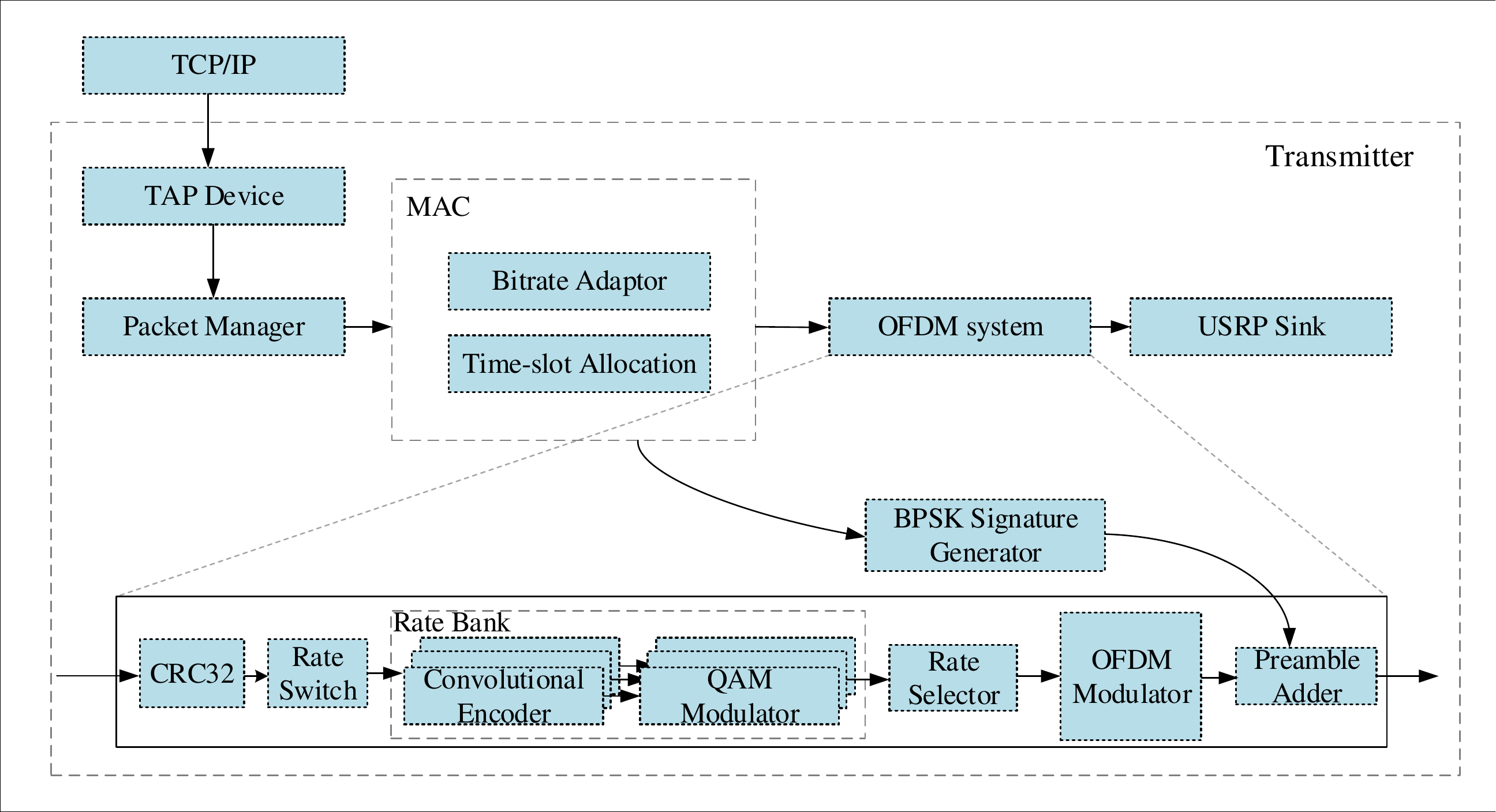}}
\caption{Transmitted-side PHY Design.}\label{fig:TXpath}
\end{figure}

\begin{figure}
\centering \scalebox{0.34}{\includegraphics{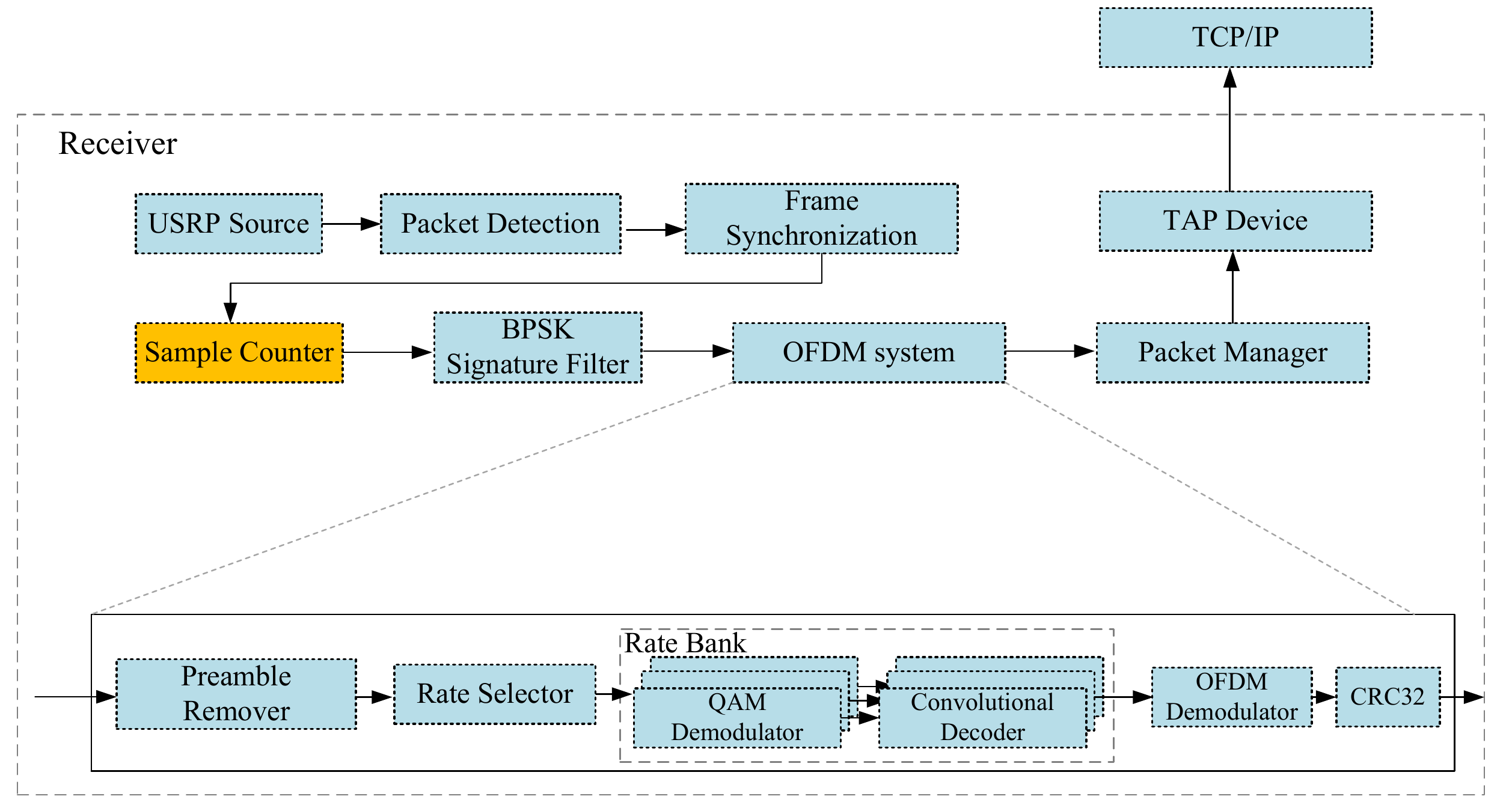}}
\caption{Receiver-side PHY Design.}\label{fig:RXpath}
\end{figure}

Fig. \ref{fig:TXpath} shows the diagram of the PHY layer at the transmitter side. The Ethernet frame from TAP is first passed to a packet manager responsible for MAC-layer operations. It computes and generates information related to rate decision, synchronization, and packet queueing. The computed results and decisions are put in tags\footnotemark{} padded along with the information bits, which are then forwarded to an OFDM system. A block, called Rate Bank, contains a bank of convolutional encoders and QAM modulators to provide bitrate variation support. Each set of a convolutional encoder and a QAM modulator gives one rate. After the QAM modulation, the symbols are then put into an OFDM modulator which performs subcarrier mapping and IFFT operations. Finally, a preamble is prepended to the packet. At the receiver side, a reverse process is performed on the received samples, as shown in Fig. \ref{fig:RXpath}.
\footnotetext{Tag, also known as Stream Tag \cite{gnuradio_streamtag}, is a mechanism in GNURadio to pass control information between blocks. Several tags can piggyback on each baseband sample. GNURadio also uses Tag to control hardware. When the samples with hardware-control tags arrive at the UHD (the USRP Hardware Driver \cite{uhd_website}), the UHD configures the USRP hardware according to the tag instructions.}

\subsection{Problem description}
RTTS-SDR is a time-slotted system which divides the channel resources into multiple time slots (see Fig. \ref{fig:timeslot_map}). In this paper, we use the upper-case $T$ to denote the time counter variable at a node. A particular value sampled from the time counter is denoted by a lower-case $t$. For example, by $T=t$, we mean the value associated with time variable $T$ is $t$ at a particular moment in time. 

Three types of local times can be kept at a node: (i) USRP time, (ii) PC time, and (iii) USRP time on PC.

\textbf{USRP time} refers to the time maintained by the time counter on the USRP. USRP time increments according to the local oscillator within the USRP. We denote the USRP time variable of  node $i$ by ${T_{{\rm{U}},i}}$ and the USRP time value by ${t_{{\rm{U}},i}}$. USRP time is purely a hardware time. It is used, for example, to time the transmission of packets by the RF board. For a node $i$, a packet is said to be transmitted at time ${t_{{\rm{U}},i}}$ (or with timestamp ${t_{{\rm{U}},i}}$) if it is transmitted when the USRP time variable ${T_{{\rm{U}},i}} = {t_{{\rm{U}},i}}$.

\textbf{PC time} is the time maintained by the PC's OS. PC time can be obtained by calling the function \textit{time.time()} within a program. We denote the PC time variable (value) in node $i$'s PC by ${T_{{\rm{P}},i}}$ (${t_{{\rm{P}},i}}$).

\textbf{USRP time on PC} is the USRP time known by the connected PC and it is denoted by ${T_{{\rm{UP}},i}}$ (${t_{{\rm{UP}},i}}$ for a value) for node $i$. Every time a program on the PC tries to get\footnotemark{}
\footnotetext{UHD provides a function \textit{get\_usrp\_hardware\_time()} for the PC to get the USRP time.}
 the \textit{USRP time} through UHD, UHD will instruct the USRP to capture the current USRP time ${T_{{\rm{U}},i}} = {t_{{\rm{U}},i}}$ and send that time to the PC. The actual USRP time is always larger than the USRP time on PC. Denote the delay\footnotemark{} between USRP and PC at node $i$ by ${\delta _i}$, we have
\footnotetext{The overall delay between USRP and PC includes the PC command processing time, delay caused by Ethernet packet transmission and USRP command response time. From our experience, the overall delay is typically dominated by the Ethernet packet delivery delay.}
\begin{equation}	\label{eqa:usrp_time_pc}
{T_{{\rm{U}},i}} = {T_{{\rm{UP}},i}} + {\delta _i}
\end{equation}

For the time-slotted mechanism, let $s_j^k$ be the \textit{global time} slot boundaries for slot $j$ of the $k$-th frame and $s_{i,j}^k$ be the time at which \textit{node $i$ thinks} slot $j$ of the $k$-th frame begins (also known as time slot boundary), i.e., $s_{i,j}^k$ is the transmission time for a packet to be transmitted in time slot $j$ of the  $k$-th frame by node $i$. If the first sample of a packet is tagged with a timestamp $s_{i,j}^k$, it will be transmitted by node $i$'s USRP when ${T_{{\rm{U}},i}} = s_{i,j}^k$. In our system, the AP's USRP time is regarded as the global reference time, that is, ${\rm{s}}_j^k = s_{0,j}^k$. Meanwhile, the time that node $i$ receives a packet from another node in slot $j$ in $k$-th frame is denoted by ${\rm{\tilde s}}_{i.j}^k$. 

The boundaries of time slots maintained by different nodes need to be aligned so as to synchronize the access of the wireless medium. As such, we need to design a synchronization mechanism tailored for the SDR platform. Our synchronization mechanism design faces two challenges: The first challenge is to align the slot boundaries of the nodes (i.e., $s_{i,j}^k$), subject to the random delays between the USRP and the PC of the nodes, different propagation delays between different pairs of nodes, and different local clock shifts. The second challenge is to ensure the USRP of a node transmits a packet precisely in a near-future time slot specified by its PC, again subject to the random delay between the PC (where the samples and instructions are generated) and the USRP (where the samples are transmitted).

To address the first challenge, we adopt a beacon synchronization and sample counting mechanism, as elaborated in Section \ref{sec:sync}. To address the second challenge, we propose an algorithm called {\textit{Just-in-time}} to ensure that packets sent out by the PC to the USRP can reach the USRP before the time slots they are to be transmitted, as elaborated in Section \ref{sec:low_latency}. \textcolor{black}{A table of notations is given in Table \ref{table:notations}.}

\begin{table}[]
	\caption{\textcolor{black}{Table of notations}}\label{table:notations}
	\centering
	{\color{black}\begin{tabular}{ m{0.15\linewidth} | m{0.75\linewidth}}
			\hline
			$i$                           &   Index of a device   \\ \hline
			$T$                         &   Time counter variable of a node \\ \hline
			$t$                          &   Value sampled from the time counter variable    \\ \hline
			${T_{{\rm{U}},i}}$          &   USRP time variable of node $i$  \\ \hline
			${t_{{\rm{U}},i}}$           &   USRP time value of node $i$ \\ \hline
			${T_{{\rm{P}},i}}$          &   PC time variable of node $i$    \\ \hline
			${t_{{\rm{P}},i}}$           &   PC time value of node $i$   \\ \hline
			${T_{{\rm{UP}},i}}$       &   Variable of USRP time on PC of node $i$ \\ \hline
			${t_{{\rm{UP}},i}}$        &   Value of USRP time on PC of node $i$    \\ \hline
			${\delta _i}$               	   &   Delay between USRP and PC at node $i$   \\ \hline
			$j$                          &   Index of the time slot  \\ \hline
			$k$                         &   Index of the time frame \\ \hline
			$s_j^k$                  &   Global time slot boundary for the beginning of slot $j$ of frame $k$ according to ${T_{{\rm{U}},0}}$    \\ \hline
			$s_{i,j}^k$                 &   Time (according to ${T_{{\rm{U}},i}}$) at which node $i$ thinks slot $j$ of frame $k$ begins    \\ \hline
			$\tilde s_{i,j}^k$          &   Time (according to  ${T_{{\rm{U}},i}}$) at which node $i$ receives a packet in slot $j$ of frame $k$    \\ \hline
			${t_{{\rm{init}},i}}$         &   Timestamp (according to ${T_{{\rm{U}},i}}$) of the first received sample of node $i$    \\ \hline
			$T_s$                       &   Duration of one time slot   \\ \hline
			$N$                         &   Size of a frame (number of time slots)   \\ \hline
			$T_{\rm{sample}}$           &   Duration of one sample   \\ \hline
			$B$                         &   Bandwidth   \\ \hline
			$t_{{\rm{beacon}},i}^k$  &   Time (according to ${T_{{\rm{U}},i}}$) of the beacon's (first) slot boundary in frame $k$ of node $i$   \\ \hline
			$\ell _i^k$                         &   Number of past samples that node $i$'s Sample Counter has counted when the beginning of frame $k$'s beacon is detected  \\ \hline
			$o_i$                       &   Clock offset between the AP's USRP clock and node $i$'s USRP clock   \\ \hline
			$d_{0,i}$                   &   Propagation delay from the AP to node $i$ \\ \hline
			$d_{i,0}$                   &   Propagation delay from the node $i$ to the AP \\ \hline
			$t_i^{(m)}$                 &   USRP time of node $i$ when Event $m$ happens  \\ \hline
			${\hat T_{{\rm{U}},0,i}}$   &   Estimation of AP's USRP time at node $i$  \\ \hline
			${\delta _{{\rm{RX}},i}}$   &   RX delay between PC and USRP in node $i$  \\ \hline
			${\delta _{{\rm{TX}},i}}$   &   TX delay between PC and USRP in node $i$  \\ \hline
			$T_{\rm{frame}}$            &   Duration of one frame   \\ \hline
			${\delta _{{\rm{RTT}},i}}$  &   Estimated Round-Trip Time (RTT) between the PC and the USRP of node $i$    \\ \hline
			${\omega _i}$               &   Safety margin for node $i$ transmits packets in advanced   \\ \hline
	\end{tabular}}
\end{table}


\section{Time Synchronization}	\label{sec:sync}
In RTTS-SDR, different nodes have different USRP times and PC times. To synchronize their times and align their time slot boundaries, a mechanism for the exchange of time information is needed so that they can reach a consensus on the time slot boundaries. We put forth a new beacon synchronization mechanism tailored for SDR.

The misalignment of the slot boundaries of different nodes can be traced for the following causes: (1) different nodes use different USRP clocks to time packet transmission and reception; (2) random delays between PC and USRP; (3) different propagation delays and clock offsets among different pairs of nodes. 

Our approach to synchronizing the slot boundaries of different nodes include: (1) a beacon synchronization mechanism to let all the IoT devices obtain information on the global reference time (i.e., AP's USRP Time), as presented in Section \ref{subsec:beacon_sync}; (2) a sample counting algorithm for a node to acquire the precise packet arrival time, as presented in Section \ref{subsec:acq_precise_time}; (3) a modified version of the Precision Time Protocol (PTP) to compensate for the propagation delays and clock offsets between AP and IoT devices, as presented in Section \ref{subsec:propagation_clockoff}. Section \ref{subsec:packet_detection} presents our algorithm for detecting the starting point of a packet. Finally, Section \ref{subsec:implication_usrp_time} explains how to leverage the accurate synchronization among the USRPs to provide an \textit{\textbf{Event Synchronization}} service to applications.

\subsection{Beacon synchronization}	\label{subsec:beacon_sync}
\subsubsection{Beacons as a time reference}

\begin{figure}
\centering \scalebox{0.42}{\includegraphics{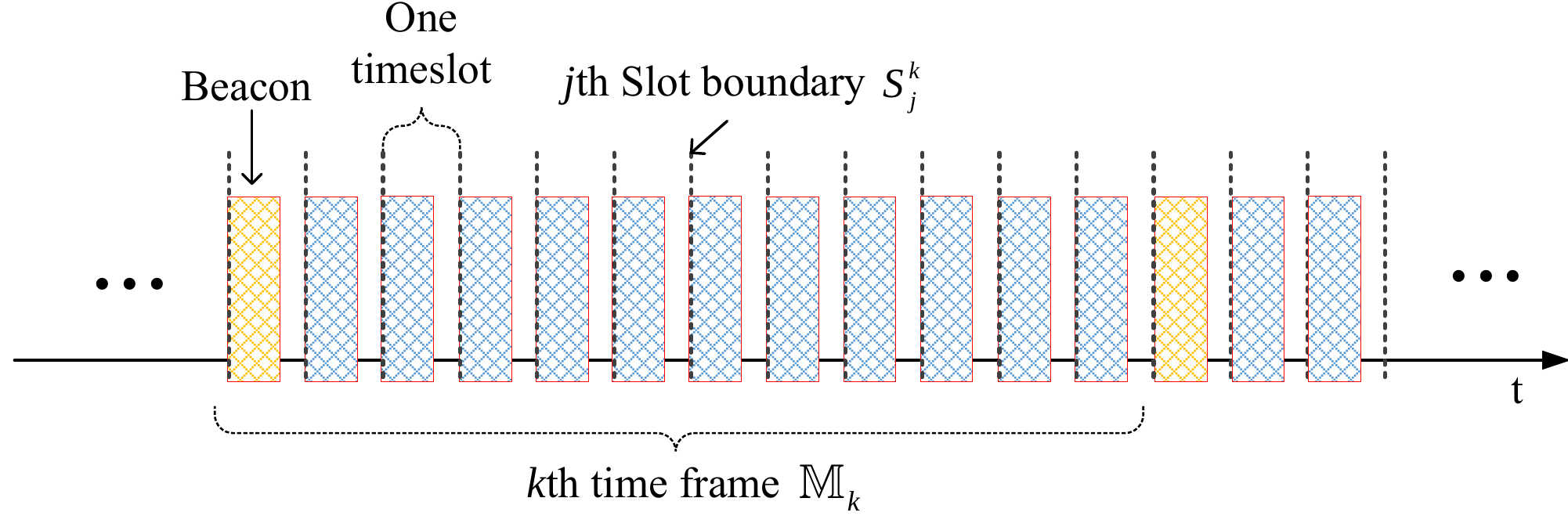}}
\caption{In RTTS-SDR, the channel resource is divided into slots and $N$ slots are grouped into a time frame.}\label{fig:timeslot_map}
\end{figure}

As shown in Fig. \ref{fig:timeslot_map}, our time-slotted system divides the channel resources into multiple time slots. A group of $N$ time slots forms a time frame. The first time slot in a time frame is dedicated to the transmission of a beacon by the AP. Beacons, used to broadcast control and feedback information, also serves as a reference packet providing timing to align the slot boundaries of the other nodes. The transmission of the beacon is timed according to the AP's USRP clock.

\subsubsection{Two phases of synchronization}	\label{subsubsec:beacon_algo}
The beacon synchronization mechanism in our system is divided into two phases: (1) beacon broadcast by the AP and (2) slot alignment by IoT devices.

In the \textbf{beacon broadcast phase}, a beacon for the $k$-th frame is first generated at the AP's PC. The AP's PC tags the beacon with a USRP transmission timestamp $s_{0,0}^k$ and then sends the beacon to the AP's USRP, which transmits it when ${T_{{\rm{U,0}}}} = s_{0,0}^k$, where ${T_{{\rm{U,0}}}}$ is the AP's \textit{USRP time}. We emphasize that the PC needs to send the beacon to the USRP ahead of ${T_{{\rm{U,0}}}} = s_{0,0}^k$ due to the random delays between the PC and the USRP. \textcolor{black}{Section \ref{sec:low_latency} puts forth a \textit{Just-in-time} algorithm to handle the random delay between the USRP and the PC.}

In the \textbf{slot alignment phase}, upon receiving the beacon, an IoT device adjusts its slot boundaries to align with the slot boundaries of the AP. Specifically, an IoT device $i$ records the beacon's arrival time according to its own USRP timer, ${T_{{\rm{U}},i}} = \tilde s_{i,0}^k$. The IoT device $i$ then computes the times of its subsequent slot boundaries, $s_{i,j}^l$, $j > 0,l \ge k$, based on $\tilde s_{i,0}^k$, as elaborated below.

\subsubsection{Time slot boundary computation}	\label{subsubsec:timeslot_compute}
For a frame of size $N$, there are $N-1$ slots following the beacon. Since we have the arrival time of the beacon $\tilde s_{i,0}^k$ in the $k$-th time frame, we can compute the time of the $j$-th time slot boundary in the $l$-th time frame $s_{i,j}^l$ by
\begin{equation}	\label{eqa:timeslot_compute}
s_{i,j}^l = \tilde s_{i,0}^k + (l - k)N{T_s} + j{T_s},
\end{equation}
where $T_s$ is the duration of a time slot (i.e., the gap between two consecutive time slot boundaries). Eq. (\ref{eqa:timeslot_compute}) shows that one can compute all the times of the slot boundaries following the beacon. We will shortly show in Section \ref{subsec:exp_accuracy} that, due to the clock drifts between the AP and the IoT devices (i.e., the clocks at different nodes may tick at slightly different rates), the slot boundary times computed by (\ref{eqa:timeslot_compute}) are no longer reliable after a few frames. Resynchronization based on a new beacon is necessary. In addition, (\ref{eqa:timeslot_compute}) has not taken the propagation delays between the AP and the IoT device into account. To do so, it will be replaced by (\ref{eqa:slot_boundary_comp_adjust}) later.

Although the idea of a two-phase synchronization mechanism seems straightforward, implementing the mechanism on the SDR platform is not trivial. In particular, it is not straightforward for the PC to acquire the exact arrival time of the beacon on its associated USRP due to the random delays between them. The beacon arrival time is essential for the GNURadio running on the PC to align slot boundaries. A simple method is to get the \textit{USRP Time} only after the beacon is decoded by the PC. However, the \textit{USRP Time on PC} is always outdated (see (\ref{eqa:usrp_time_pc})). Next, we elaborate on our method to acquire the precise beacon arrival time. 

\subsection{Acquisition of precise arrival time}	\label{subsec:acq_precise_time}
To acquire the precise beacon arrival time, we need to circumvent the effects of the inter USRP-PC delay and the PC beacon decoding delay. It turns out that although the USRP does not provide the arrival time of a packet or a sample directly, it gives the timestamp ${t_{{\rm{init}},i}}$ of the USRP time counter when its RX path is first started up by the PC during initialization. This timestamp is then piggybacked on the first received sample in the RX path and sent to the PC. In other words, the PC has the \textit{USRP time} when the first sample is received on the RX path. We can derive the \textit{USRP time} of the later samples by a sample counting process, as elaborated below.

In our system, the bandwidth is $B$, and thus the duration of one sample ${T_{{\rm{sample}}}}$ is $1/B$. As shown in Fig. \ref{fig:RXpath}, we add a block called \textbf{Sample Counter} after the \textit{Frame synchronization} block. When the USRP receiver outputs new samples (Note that the USRP receiver samples signal from the air channel continuously without interruption), the \textit{Sample Counter} counts the number of incoming samples. Once the Frame synchronization block detects a peak signifying the beginning of a beacon, it attaches a special tag to the sample corresponding to the beginning of the beacon. \textcolor{black}{The \textit{Sample Counter}, upon detecting the special tag, then computes the tagged sample's \textit{USRP Time} by
\begin{equation}	\label{peak_compute}
 t_{{\rm{beacon}},i}^k = {t_{{\rm{init}},i}} + {\ell _i^k} \cdot {T_{{\rm{sample}}}},
\end{equation}
where ${\ell _i^k}$ is the number of samples that node $i$'s \textit{Sample Counter} has counted (since system initialization) when the tagged sample is detected. 
The time value of this special tagged sample is then used as the arrival time of the beacon (which is also the time of the first slot boundary) in frame $k$:
\begin{equation}
\tilde s_{i,0}^k = t_{{\rm{beacon}},i}^k.
\end{equation}
}

We remark that the expression in (\ref{eqa:timeslot_compute}) is the boundary of a time slot computed based on the beacon received at the receiver of IoT device $i$. In general, different IoT devices will have different slot boundaries because of the different propagation delays from the AP to the IoT devices. Also, the packets transmitted by them to the AP may incur different propagation delays. Our goal is to synchronize and align the slot boundaries of different devices as perceived by the receiver of the AP. Toward that end, we will first need to estimate the propagation delay. The propagation delay can be estimated by applying a modified version of PTP \cite{ptp_ieee}.

\subsection{Propagation delay and clock offset estimation}	\label{subsec:propagation_clockoff}
Let ${o_i}$ be the clock offset between the AP's USRP clock and an IoT device $i$'s USRP clock, ${d_{0,i}}$ be the propagation delay from the AP to the IoT device $i$, and ${d_{i,0}}$ be the propagation delay in the opposite direction. Suppose the AP sends the $k$-th beacon at its USRP time ${T_{{\rm{U,0}}}} = s_{0,0}^k$. The AP records that time into the beacon's payload as timing information to convey the IoT devices. When the beacon is transmitted at ${T_{{\rm{U,0}}}} = s_{0,0}^k$, the USRP time at the IoT device $i$ is ${T_{{\rm{U}},i}} = s_{0,0}^k + {o_i}$. When the beacon arrives at the IoT device's receiver side (event 1), the USRP time of the IoT device is
\begin{equation}	\label{eqa:event_1_i}
t_i^{(1)} = \tilde s_{i,0}^k = s_{0,0}^k + {o_i} + {d_{0,i}}.
\end{equation}
At this moment, the USRP time of the AP is $t_0^{(1)} = s_{0,0}^k + {d_{0,i}}$. Note that  the IoT device $i$ can retrieve the beacon transmission time $s_{0,0}^k$ at the AP from the beacon payload, and it can obtain the value of ${t_i^{(1)}}$ according to its own USRP clock, ${T_{{\rm{U}},i}}$. To the IoT device $i$, the unknowns in (\ref{eqa:event_1_i}) at this point are ${d_{0,i}}$ and $o_i$, which are to be estimated.

\begin{figure}
\centering \scalebox{0.67}{\includegraphics{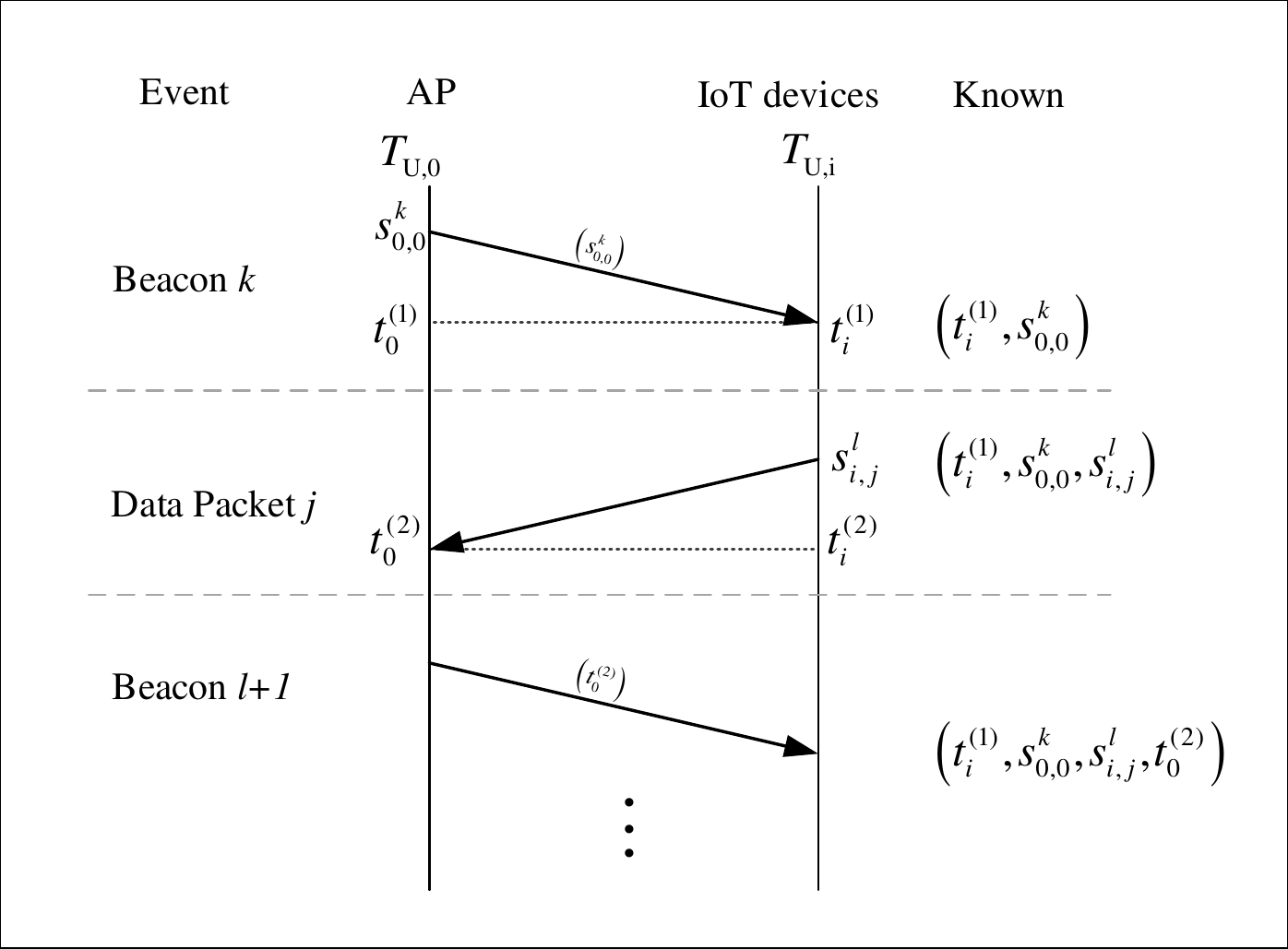}}
\caption{A three-way handshake scheme based on PTP for both clock offset compensation and propagation delay compensation.}\label{fig:ptp_scheme}
\end{figure}

In a near-future time slot $j$ of frame $l$ allocated to the IoT device $i$, the IoT device $i$ sends a data packet at its USRP time $s_{i,j}^l$, and $s_{i,j}^l$ is recorded into the payload of the packet. When the AP receives the corresponding data packet (event 2), its USRP time is 
\begin{equation}	\label{eqa:event_2_0}
t_0^{(2)} = s_{i,j}^l + {d_{i,0}} - {o_i}.
\end{equation}
At this moment, the USRP time of the IoT device $i$ is $t_i^{(2)} = s_{i,j}^l + {d_{i,0}}$. In the next transmitted beacon, the AP embeds $t_0^{(2)}$ into the beacon's payload. By the time the IoT device receives the new beacon, it knows four values: $t_i^{(1)},s_{0,0}^k,s_{i,j}^l,t_0^{(2)}$. Therefore, it can estimate the propagation delay $d_{0,i}$ by combining (\ref{eqa:event_1_i}) and (\ref{eqa:event_2_0}):
\begin{equation}	\label{eqa:delay_estimation}
{d_{0,i}} = {d_{i,0}} = \frac{1}{2}\left( {t_i^{(1)} + t_0^{(2)} - s_{0,0}^k - s_{i,j}^l} \right),
\end{equation}
\begin{equation}	\label{eqa:clock_offset_compute}
{o_i} = \frac{1}{2}\left( {t_i^{(1)} - t_0^{(2)} - s_{0,0}^k + s_{i,j}^l} \right),
\end{equation}
assuming that the propagation delays of both directions are the same (i.e. ${d_{0,i}} = {d_{i,0}}$) and the clock offset $o_i$ is constant. In Section \ref{subsec:exp_accuracy} we will show that the clock offset $o_i$ remains constant for a duration that is much longer than one frame duration. A diagram in Fig. \ref{fig:ptp_scheme} shows the exchange and the acquisition of the timing information between the AP and the IoT device $i$.

The relationship between $s_{0,0}^k$ and $\tilde s_{i,0}^k$ can be written as
\begin{equation}	\label{eqa:slot_boundary_relationship}
\tilde s_{i,0}^k = s_{0,0}^k + {d_{0,i}} + {o_i}.
\end{equation}
Since the propagation delay is well estimated, to align the arrival of packets transmitted by different IoT devices at the AP, the transmission time of a packet for the $j$-th time slot in $l$-th time frame of the IoT device $i$ should be set to
\begin{equation}	\label{eqa:slot_boundary_comp_adjust}
s_{i,j}^l = \tilde s_{i,0}^k + (l - k)N{T_s} + j{T_s} - 2{d_{0,i}}.
\end{equation}
Note that (\ref{eqa:slot_boundary_comp_adjust}) is different from (\ref{eqa:timeslot_compute}) in that it lets the IoT device transmit its packet $2{d_{0,i}}$ (i.e. round-trip delay) earlier. If all IoT devices do this, then the transmission boundaries of the AP and the reception boundaries associated with all IoT devices align at the AP.

\subsection{Packet detection}	\label{subsec:packet_detection}
As discussed in Section \ref{subsubsec:beacon_algo}, to find the precise arrival time of the beacon/packet, the \textit{Frame Synchronization} block must be able to find the first sample of a beacon correctly. Packet detection (finding the beginning of a packet from a train of received samples) is a common issue in asynchronous wireless communication networks in which nodes can generate and transmit packets at arbitrary times. There are two reasons why we still need packet detection in our synchronous time-slotted system: (i) our OFDM system is modified from that of a WiFi system, which is asynchronous in operation; (ii) finding the positions of beacons is still important for the purpose of slot alignment even if we do not use \textit{Frame Synchronization} to detect regular packets.

There are many methods for packet detection. Our method is based on a modification of the method in \cite{Bloessl2013}. We refer the interested readers to Sections 2.2 and 2.4 of \cite{Bloessl2013} for details. We modified the detection algorithm to provide the index of the first sample of a packet and piggyback a special tag on that sample. The \textit{Sample Counter} block can then extract the special tag and determine the time of the future time slot boundaries. Specifically, $\tilde s_{i,0}^k$ in (\ref{eqa:slot_boundary_comp_adjust}) can be expressed as

\begin{equation}
	\tilde s_{i,0}^k = {t_{{\rm{init}},i}} + {I_{{\rm{start}},i}^k} \cdot {T_{{\rm{sample}}}},
\end{equation}
where ${I_{{\rm{start}},i}^k}$ is the index of the first sample of the beacon received by the IoT device $i$ in the $k$-th frame.


\subsection{Implications of synchronized USRP times}	\label{subsec:implication_usrp_time}
By synchronizing the time-slot boundaries among the AP and the IoT devices, one can further synchronize the USRP times among the nodes. Specifically, the offset between the AP's USRP clock and the IoT device $i$'s USRP clock, $o_i$, is obtained in (\ref{eqa:clock_offset_compute}) at the IoT device $i$. The IoT device $i$ can estimate the AP's USRP time ${\hat T_{{\rm{U}},0,i}}$ by
\begin{equation}	\label{eqa:usrp_time_implication}
{\hat T_{{\rm{U}},0,i}} = {T_{{\rm{U}},i}} + {o_i}.
\end{equation}
By having the precise time of the global clock in every IoT device, a service that relies on the timing information of the global clock can be provided to other applications besides just our time-slot alignment application, with the global time being the AP time. We name the service as \textit{\textbf{Event Synchronization}}. For example, if we want different IoT devices to perform synchronized actions at a particular point in time, this service can be used to make sure these actions are indeed performed according to a common time.

With this service, applications that require microsecond-level synchronization can now be handled. For an event $E$ to happen at global time $t_{E}$ at the IoT device $i$ can be scheduled to happen at local time ${T_{{\rm{U}},i}} = {t_E} + {o_i}$. For example, if the IoT devices are sensors, we could coordinate them to make a measurement at exactly the global time $t_{E}$ in a synchronized manner.

{\color{black}
\begin{remark}
If a beacon is not detected due to noise or interference, the IoT device's future slot boundaries are not updated. A missed beacon, however, is not a big concern because of the high accuracy of the USRP oscillators ($2.5$ ppm). The slots of different IoT devices will not drift apart by more than a sample if the beacons of a small number of consecutive frames are missed.
\end{remark}

\begin{remark}
The TDMA system running with our synchronization mechanism can support a large number of IoT devices in principle. However, for a practical system, the maximum number of supported devices is constrained by the required minimum cycle time---the time needed by the AP to exchange one packet with every IoT device \cite{luvisotto2017physical}.
\end{remark}
}

\section{Low Latency Transmission}	\label{sec:low_latency}
This section develops a packet transmission scheme, named {\textit{Just-in-time}}, to reduce delays in packet transmission. The preparation of a packet to be transmitted consists of two steps: (1) preparation of the baseband samples; (2) preparation of the USRP timestamp for the first baseband sample's tag. Recall that a packet to be transmitted at \textit{USRP Time} $t_{{\rm{U}},i}$ needs to be tagged with a timestamp $t_{{\rm{U}},i}$. When a packet with a timestamp $t_{{\rm{U}},i}$ is sent to the USRP, it will be put in a \textit{SampleQueue} \cite{usrp_sync} in the USRP to wait for transmission at time ${T_{{\rm{U}},i}} = t_{{\rm{U}},i}$. If the USRP finds out that its current hardware time ${T_{{\rm{U}},i}} > t_{{\rm{U}},i}$, it will drop the packet and return a status ``L'' (which stands for Late \cite{uhd_usrp_late_doc}) to the PC and the transmission is considered to have failed. Our system needs to avoid such failures. 

In our time-slotted system, packet transmission faces two challenging issues:
\begin{enumerate}[(a)]
\item The PC does not have direct access to the current \textit{USRP Time}.	\label{enum:issue_1}
\item There is an uncontrollable delay between the PC and the USRP.	\label{enum:issue_2}
\end{enumerate}
In other words, the PC needs to estimate the \textit{USRP Time} indirectly and prepare the packet in advance to compensate for the delay.

For challenge (a), although the \textit{Sample Counter} does not provide the current \textit{USRP Time}, it provides the most updated \textit{USRP Time on PC}. When the PC calls a function provided by the \textit{Sample Counter} to get the \textit{USRP Time}, it returns the latest number of samples that have passed through it. In other words, the value returned by the \textit{Sample Counter} represents the time of the latest sample coming from the USRP, which only experiences the delay between the PC and the USRP, and the value is
\begin{equation}	\label{eqa:sample_counter_delay}
{T_{{\rm{UP}},i}} = {T_{{\rm{U}},i}} - {\delta _{{\rm{RX}},i}},
\end{equation}
where ${\delta _{RX,i}}$ is RX delay between PC and USRP in IoT device $i$.

\begin{figure}
	\centering \scalebox{0.37}{\includegraphics{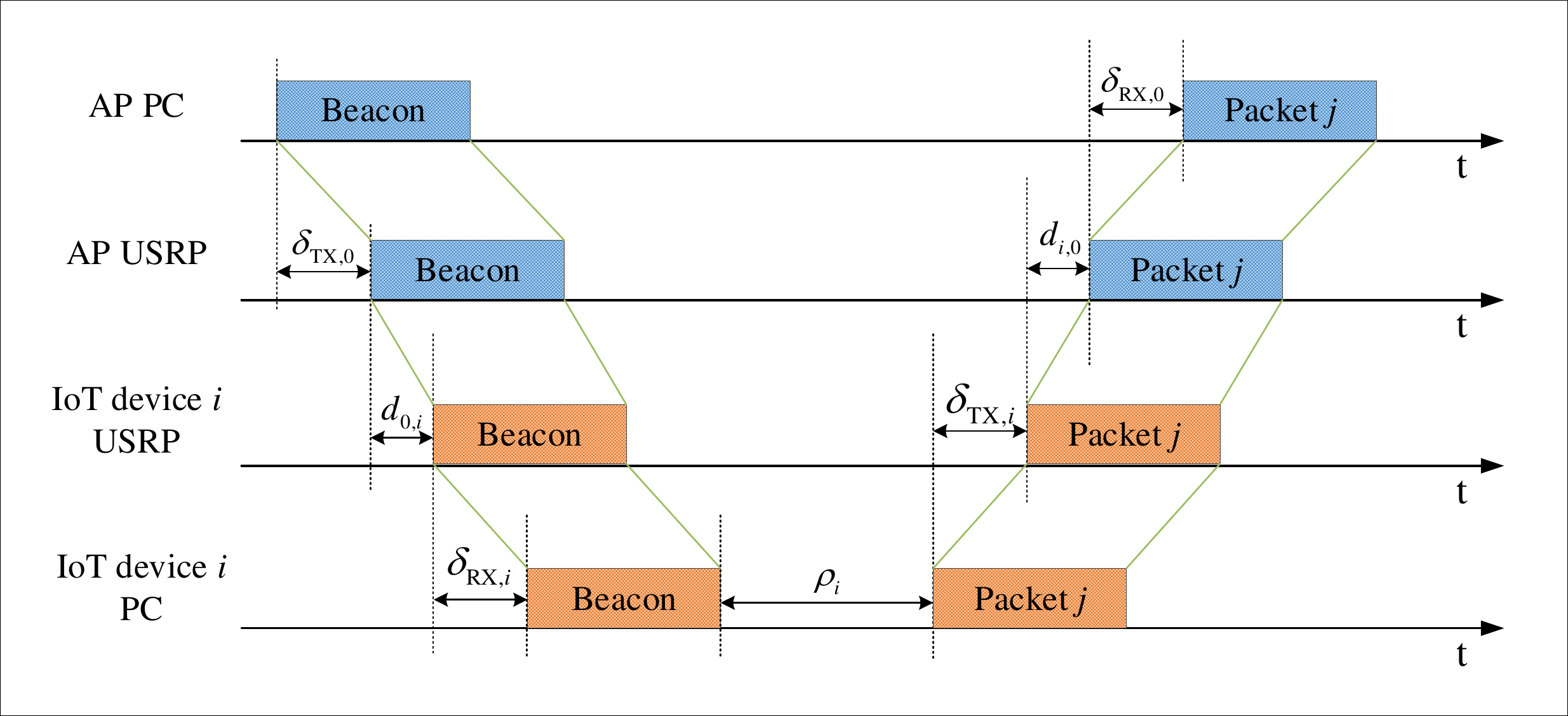}}
	\caption{\textcolor{black}{An example showing how delays affect the transmission and reception of beacons/packets.}}\label{fig:time_relation}
\end{figure}

\textcolor{black}{For challenge (b), an illustration  is shown in Fig. \ref{fig:time_relation}, where ${\delta _{{\rm{TX}},0}}$ (${\delta _{{\rm{TX}},i}}$) is the PC-USRP delay at the AP (IoT device $i$), and ${\delta _{{\rm{RX}},0}}$ (${\delta _{{\rm{RX}},i}}$) is the delay at the reverse direction at the AP (IoT device $i$).\footnotemark{} When the AP's PC sends a beacon, it takes ${\delta _{{\rm{TX}},0}}$ amount of time for it to arrive at the AP's USRP. If the USRP transmits the beacon immediately, then after ${d_{0,i}}$ propagation delay, the beacon arrives at IoT device $i$'s USRP. The USRP then sends the beacon to the PC, incurring an additional ${\delta _{{\rm{RX}},i}}$ delay. The PC takes ${\rho _i}$ amount of time to process the samples and prepare packet $j$.}

\footnotetext{We omit the USRP hardware's circuit time in this example.}

\textcolor{black}{If the PC of a node does not take the delay into consideration and sends a packet to the USRP at \textit{USRP Time} ${T_{{\rm{U}},i}} = {t_{{\rm{U}},i}}$ with the timestamp ${t_{{\rm{U}},i}}$, when the packet arrives at the USRP hardware, \textit{USRP Time} is already ${T_{{\rm{U}},i}} = {t_{{\rm{U}},i}} + {\delta _{{\rm{TX}},i}}$. Therefore, the PC of node $i$ needs to make sure the packet is sent to the \textit{SampleQueue} at least ${\delta _{{\rm{TX}},i}}$ in advance.}

A straightforward solution is to let the PC intentionally prepare the packet content and its timestamp far before the targeted transmission time. For example, when a beacon arrives at the node's PC, the node immediately prepares the packets to be transmitted $K$ time frames later where $K \gg 1$ and sends these packets to the USRP. However, this method is not viable for most time-sensitive IIoT applications. This is because if the packet is prepared far before its transmission time, say $t_{{\rm{U}},i} = {t_{{\rm{UP}},i}} + K \cdot {T_{{\rm{Frame}}}}$, where ${T_{{\rm{Frame}}}}$ is the duration of a time frame, the end-to-end delay of the packet will become excessive. For example, if this packet contains a sensor reading or a control command, the reading or command will not be fresh with this early preparation of the packet.

To achieve low latency transmission, all unnecessary overhead delay should be removed before the packet transmission. For a time-slotted multiuser system, each IoT device can only transmit its packets at its pre-allocated time slots. Taking issues (a) no direct access to \textit{USRP time} and (b) uncontrollable delay between PC and USRP into consideration, the timing for sending packets from PC to USRP needs to be carefully set. In the following, we propose an effective mechanism to deal with this issue.

\subsection{{\textit{Just-in-time}} transmission scheme} 
To prevent the PC from generating and sending a packet to the USRP too much ahead of its transmission time, we put forth a scheme called {\textit{Just-in-time}}. In essence, the PC generates and sends packets based on an estimated transmission delay. For a packet to be transmitted in time slot $j$ of frame $k$ by node $i$, it needs to be sent by the PC at
\begin{equation}	\label{eqa:wake_ahead_a_bit_algo}
{t_{{\rm{UP}},i}} = s_{i,j}^k - {\delta _{{\rm{TX}},i}},
\end{equation}
where ${\delta _{{\rm{TX}},{\rm{i}}}}$ is the delay from the PC to the USRP. Combining (\ref{eqa:sample_counter_delay}) and (\ref{eqa:wake_ahead_a_bit_algo}), we know that node $i$'s PC needs to send the packet to the USRP on or before
\begin{equation}	\label{eqa:send_in_adv_compute}
{t'_{{\rm{UP}},i}} = s_{i,j}^k - {\delta _{{\rm{TX}},i}} - {\delta _{{\rm{RX}},i}} = s_{i,j}^k - {\delta _{{\rm{RTT}},i}},
\end{equation}
where ${\delta _{{\rm{RTT}},i}}$ is the estimated round-trip time (RTT) between the PC and the USRP hardware of the $i$-th node. The question then becomes how to estimate the RTT. Fortunately, a simple tool can be used to obtain the delay between the PC and the USRP: \textit{PING test}. Internet Control Message Protocol (ICMP) packets, which are sent by the \textit{ping} command on the PC, go through the Ethernet to reach the USRP, and the USRP will give ICMP response packets back to the PC.

Because the RTT between the PC and the USRP has large jitters, safety margins need to be added. We follow how the TCP adds safety margin to handle possible jitters in setting the RTO \cite{garcia2004}. In particular, the safety margin for the  $i$-th node is calculated by
\begin{equation}	\label{eqa:safety_margin}
{\omega _i} = \beta  \cdot {\sigma _{{\rm{RTT}},i}},
\end{equation}
where $\beta$ is an adjustable coefficient and ${\sigma _{{\rm{RTT}},i}}$ is the measured deviation of the RTT of the $i$-th node. Consequentially, the packet targeted at the $j$-th time slot should be sent by the $i$-th node's PC at
\begin{equation}	\label{eqa:new_packet_tx_time}
{t''_{{\rm{UP}},i}} = s_{i,j}^k - {\delta _{{\rm{RTT}},i}} - \beta  \cdot {\sigma _{{\rm{RTT}},i}}.
\end{equation}

\textcolor{black}{From this point, the algorithm becomes simple. We define ${{\rm{T}}_{adv}}$ to be the time for the PC to start the transmission procedure in advance, and we have
\begin{equation}	\label{eqa:t_advance_definition}
{{\rm{T}}_{adv}} = {\delta _{{\rm{RTT}},i}} + \beta  \cdot {\sigma _{{\rm{RTT}},i}}.
\end{equation}}
\textcolor{black}{Since GNURadio is a thread-based program that separates the processing of packets into many threads, the packet preparation and transmission in GNURadio on the PC runs on a thread while the reception of packets runs on another thread. In that light, if node $i$ wants to transmit a packet in slot $j$ of frame $k$, we set the transmission thread to a ``ready'' state and set a countdown timer with the initial value  $s_{i,j}^k - {T_{adv}}$.}

\begin{figure}
\centering \scalebox{0.5}{\includegraphics{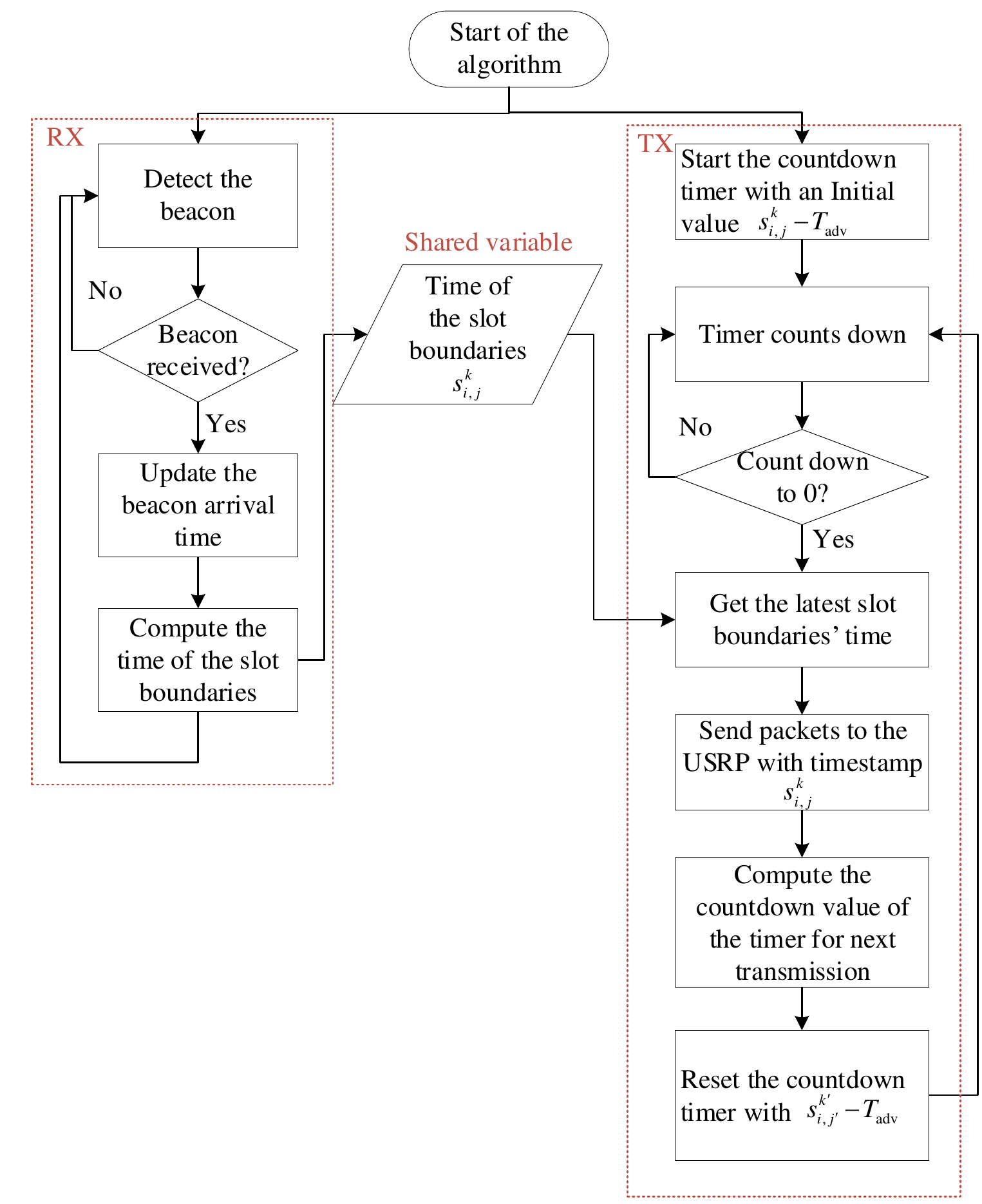}}
\caption{\textcolor{black}{The thread flow graph of the \textit{Just-in-time} algorithm.}}
\label{fig:RX_Flowchart}
\end{figure}

\textcolor{black}{When the timer counts to zero, the thread goes to the ``running'' state, whereupon it prepares the packet content, tags the timestamp $s_{i,j}^k$ to the packet, and then sends it to the \textit{SampleQueue} in the USRP. After that, the timer is reset to $s_{i,j'}^{k'} - {T_{adv}}$ again with $j' \ge j$ and $k' \ge k$, where $(j',k')$ is the slot for the next transmission. A flowchart is provided in Fig. \ref{fig:RX_Flowchart} to show the procedures of the algorithm.}

\subsection{Implication of {\textit{Just-in-time}} algorithm} 
For a wireless IoT network that leverages the {\textit{Just-in-time}} algorithm, its AP/IoT devices' applications must generate packet(s) ${T_{adv}}$ ahead of the USRP transmission time. Taking the wireless sensor network as an example, if its application is a sensor sensing data, the sensor will takes a measurement and report it at time ${T_{adv}}$ ahead of the transmission time. The return packet from the AP could be a packet generated by a controller based on the sensed data. Thus, by using the minimum acceptable $T_{adv}$, the {\textit{Just-in-time}} algorithm can reduce the round-trip delay of a feedback loop in a control system.

\section{Experimental Validation}	\label{sec:experiments}
\begin{figure}
\centering \scalebox{0.2}{\includegraphics{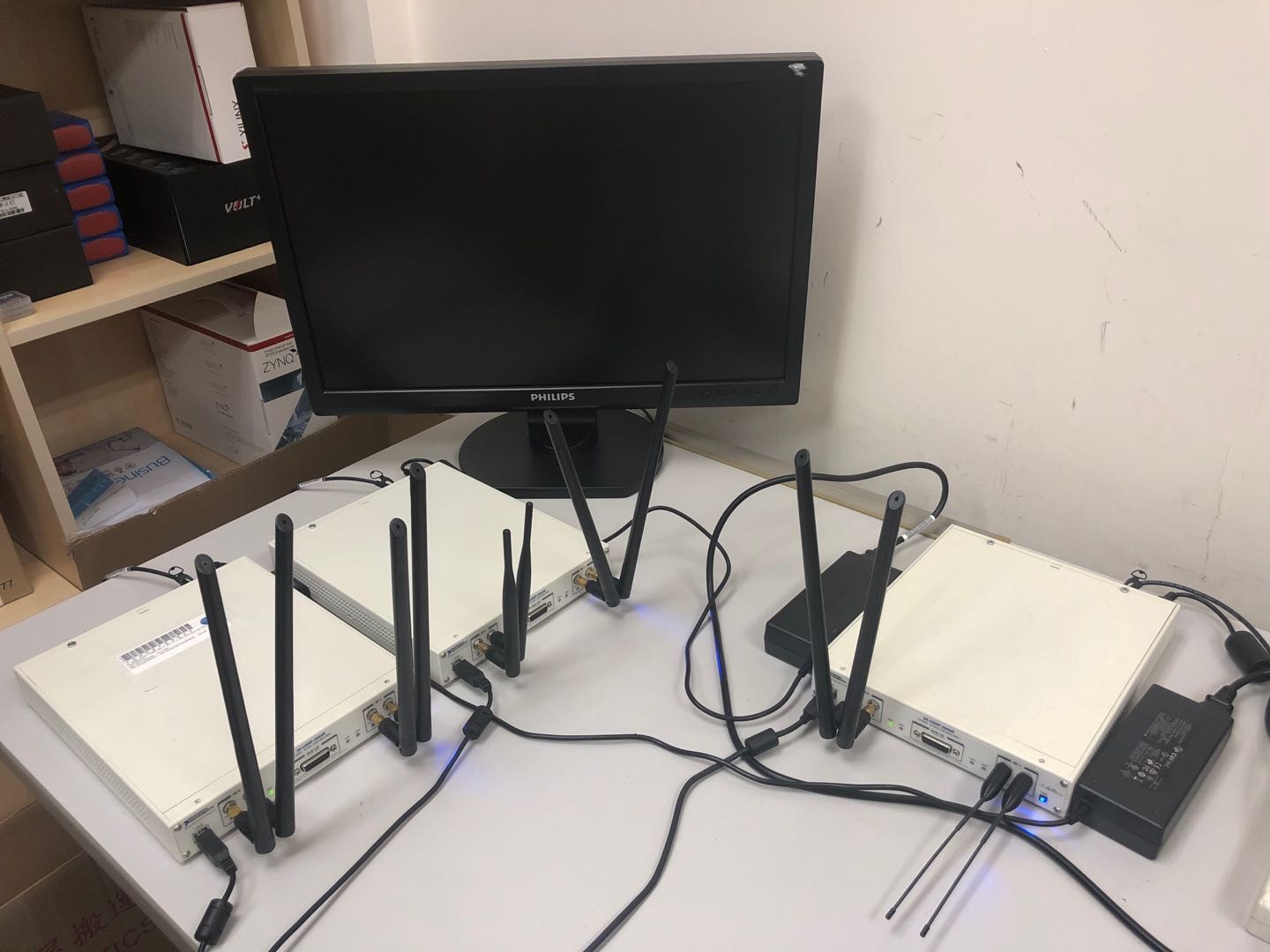}}
\caption{The testbed of our experiment. Three sets of Powerful PC + USRP in an indoor office environment. One of them serves as the AP and the other two are the IoT devices.}\label{fig:testbed}
\end{figure}

To evaluate RTTS-SDR, we deployed three sets of USRP X310 with onboard TCXO and UBX-160 daughterboards \cite{usrp_specs} (i.e., there are three nodes, one of which is the AP; see Fig. \ref{fig:testbed}). Each of the USRP is connected to a PC with a $10$Gbps Ethernet cable. The PC has a 16-core AMD 1950X Processor 3.4GHz and 64G RAM. The operating system is Ubuntu 16.04 LTS with kernel version 4.15.0-60-generic, installed with UHD 3.9.7 and GNURadio 3.7.11.

The PHY-layer adopts the settings as shown in TABLE \ref{table:exp}. The number of time slots in each time frame, including the beacon, is $19$. Three nodes transmit data packets in a round-robin manner. That is, the time slot of each of the nodes is pre-assigned. To clearly show the structure of time slots and the demarcation between two slots, we introduce $360$-sample time as the guard time between two consecutive slots. We emphasize that in actual system, this excessive guard time is not necessary. It is purely an artificial setting so that we can visually delineate the different time slots in Fig. \ref{fig:samples}.

\begin{table}[]
\caption{\textcolor{black}{Parameters of the PHY-layer in the experiment}}\label{table:exp}
\centering
{\color{black}\begin{tabular}{l|l}
\hline
Center frequency   & $2.418$GHz               \\ \hline
Bandwidth          & $10$MHz                  \\ \hline
Length of payload      & $128$ OFDM symbols       \\ \hline
Modulation         & BPSK                     \\ \hline
Channel code       & $1/2$ convolutional code \\ \hline
Length of preamble & $4$ OFDM symbols        \\ \hline
Length of cyclic-prefix (CP) & $16$ samples        \\ \hline
Guard time	& $360$ samples			\\	\hline
Long training sequence (LTS) & $80 \times 2$ samples			\\ \hline
Short training sequence (STS) & $80 \times 2$ samples 	\\ \hline
\end{tabular}}
\end{table}

\subsection{Usability of the time-slotted system}	\label{subsec:exp_usability}
We ran the system in an office environment. We captured the signal at the receiver side of the AP. The transmit power of three nodes are intentionally not carefully calibrated to better emulate practical scenarios.

\begin{figure}
\centering \scalebox{0.3}{\includegraphics{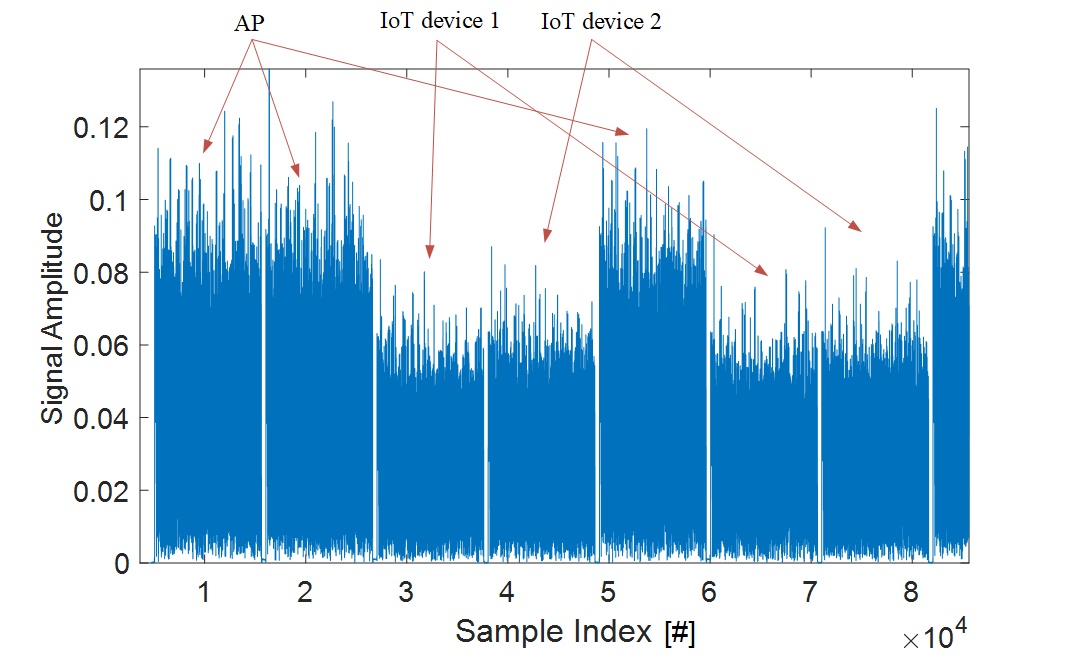}}
\caption{An example of round-robin TDMA implemented on RTTS-SDR. We can see visually that the time-slot boundaries of different nodes are synchronized.}\label{fig:samples}
\end{figure}

As shown in Fig. \ref{fig:samples}, the first time slot is a beacon, followed by round robin transmissions of data packets from all three nodes, including the AP. Note that the round-robin transmission scheme is just an example which can be changed based on the traffic requirements. The changes can also be done in real-time by the AP by embedding the scheduling changes in the beacon (i.e., beacons, besides serving as a timing reference, also contain instructions from the AP).

\subsection{Accuracy of synchronization}	\label{subsec:exp_accuracy}
We next tested the accuracy of our synchronization algorithm. We first measured the clock drift between the IoT device $1$ and the AP in the absence of synchronization. To do so, we compared the timestamps of received beacons and the expected timestamps of the same beacons at the IoT device $1$ (i.e., $i=1$). \textcolor{black}{To simplify notation, we remove the subscript $i$ in $t_{{\rm{beacon}},i}^k$. The timestamp of the received beacon for frame $k$ is $t_{\rm{beacon}}^k$, and the expected timestamp of the beacon in frame $k$ can be computed by 
\begin{equation}	\label{eqa:beacon_comp}
t_{{\rm{beacon}}}^k = t_{{\rm{beacon}}}^0 + kN{T_s}.
\end{equation}
On the other hand, the actual timestamp of the received beacon in frame $k$ is $\tilde t_{{\rm{Beacon}}}^k$. Therefore, the clock drift at frame $k$ of the IoT device $1$ can be defined as
\begin{equation}
\Delta t_{{\rm{beacon}}}^k = t_{{\rm{beacon}}}^k - \tilde t_{{\rm{beacon}}}^k.
\end{equation}
}

\begin{figure}
\centering \scalebox{0.67}{\includegraphics{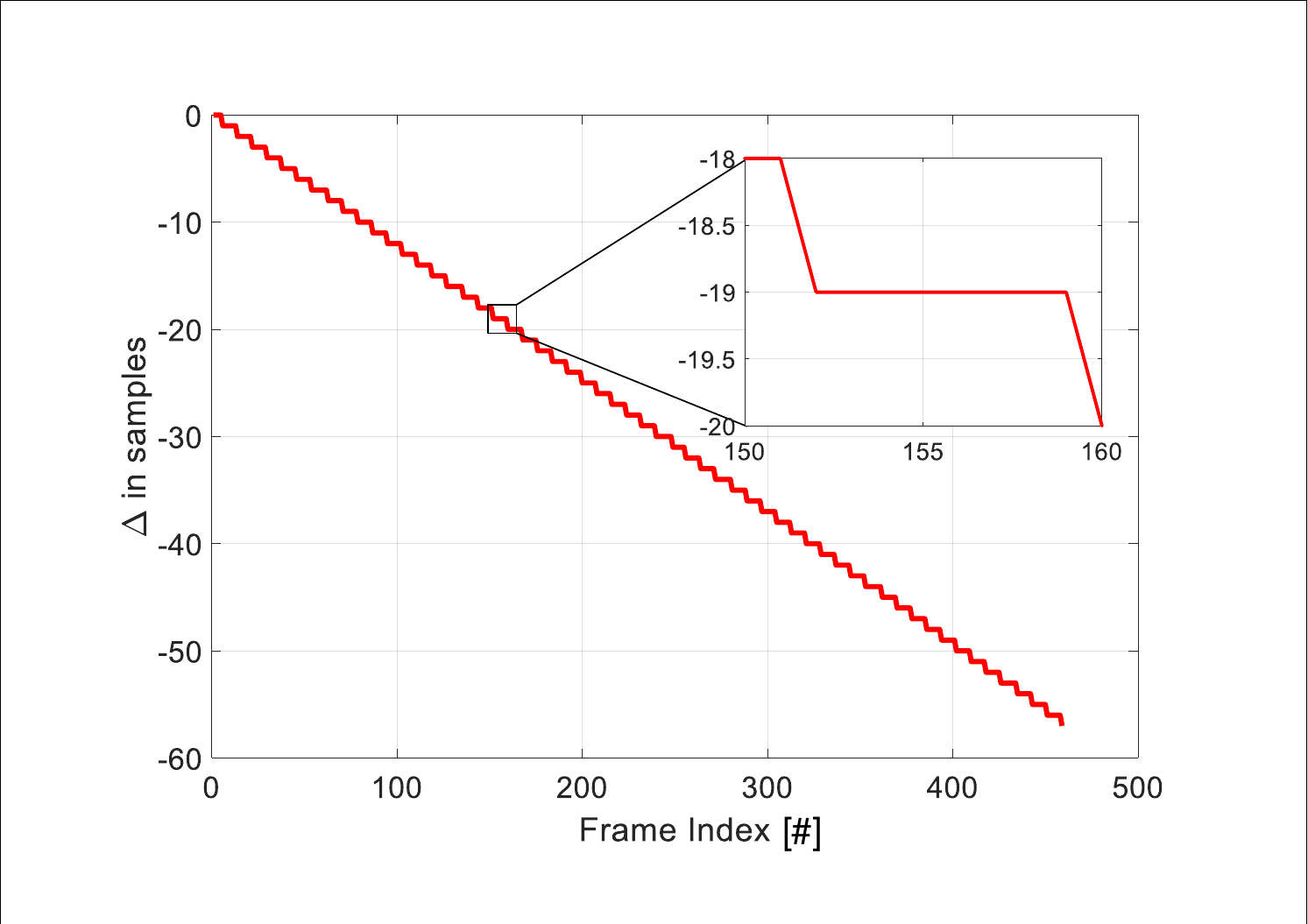}}
\caption{Clock drift $\Delta $ (in units of number samples) between the AP's USRP and one of the IoT devices' USRP.}\label{fig:clock_drift}
\end{figure}

Fig. \ref{fig:clock_drift} shows the clock drift in terms of samples. The duration of one time slot is
\begin{align}
{T_s} &= \left[ {(128 + 4) \times (64 + 16) + 360} \right] \cdot {T_{{\rm{sample}}}} \notag \\
 &= 10920 \times \frac{1}{{10 \times {{10}^6}}} = 1.092{\mkern 1mu} ms.
\end{align}
Recall that there are $19$ slots in one frame in our experiment, the duration of one frame is thus $1.092 \times 19 = 20.748{\mkern 1mu} ms$. As shown in Fig. \ref{fig:clock_drift}, the clocks of two nodes drift apart by more than $5$ samples after $1$ second (i.e., $48.197$ frames). After $400$ frames, the clock drift increases to $50$ samples. Even disregarding the propagation time and considering only the clock drift, the guard time ${T_{{\rm{Guard}}}} = 360{T_{{\rm{sample}}}}$ can only tolerate around $2880$ frames. In other words, in about one minute at most, the transmissions from two nodes in two consecutive slots will collide with each other if we do not have the time-slot alignment procedure. The curve in Fig. \ref{fig:clock_drift} is staircase-like since the clocks drift apart by less than one sample between two consecutive frames: they drift apart by one sample after a few frames. From the result, we can also conclude that when synchronization is turned on, and if an IoT device misses the detections of a few successive beacons and cannot perform synchronization for the few successive frames, the alignment of its slot boundaries will still be within one sample (about $1$ sample drift every $7$ frames in the lack of synchronization).

To measure the accuracy of our synchronization algorithm, we analyze the timing of the received signals at the AP. Specifically, since the USRP clock of the AP is used as the reference clock, and the AP's RX path provides a timestamp for each of the received packets (includes its own packets), the timestamp can be used to measure each node's synchrony. Let $\tilde R_{i,j}^k$ be the actual timestamp of the received packet sent from node $i$ in the $k$-th frame's slot $j$ and $R_{i,j}^k$ be the corresponding expected timestamp. The absolute difference of two timestamps $\Delta R_{i,j}^k = \left| {R_{i,j}^k - \tilde R_{i,j}^k} \right|$ is used as the metric to quantify the synchrony between the AP and the IoT device $i$. 

\begin{figure}
\centering \scalebox{0.5}{\includegraphics{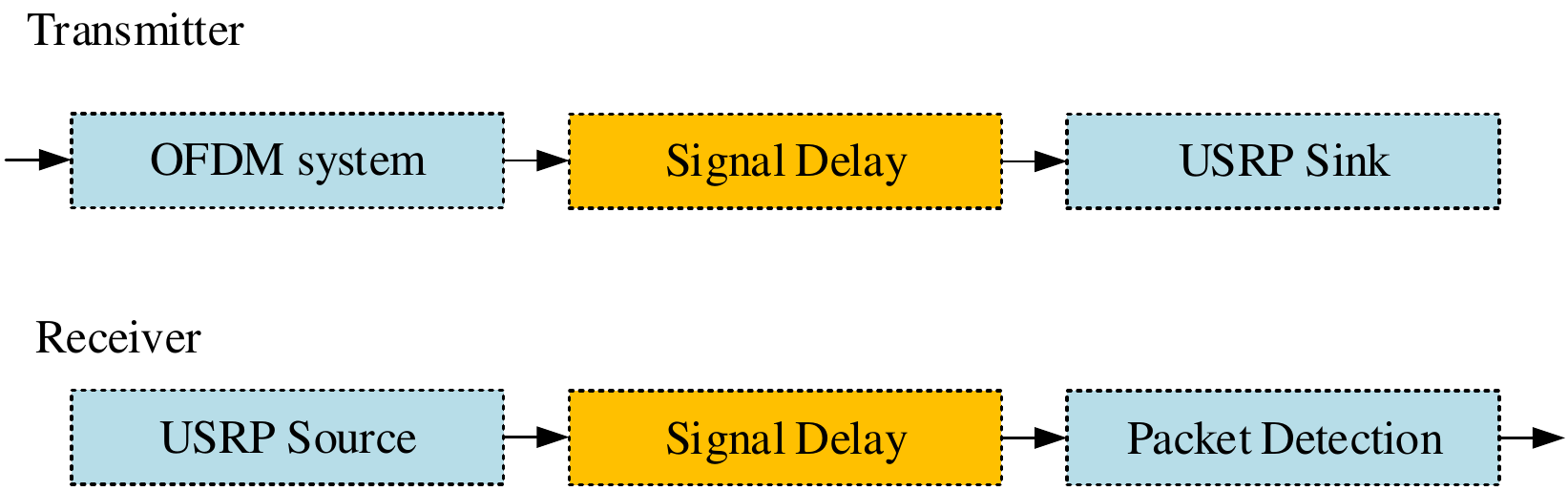}}
\caption{The locations of adding the Signal Delay block in the GNURadio TX and RX paths of the IoT devices.}\label{fig:signal_delay}
\end{figure}

To evaluate $\Delta R_{i,j}^k$ with non-negligible propagation delays, we emulated a $90$-meter propagation path between the AP and one of the IoT devices (IoT device $1$) and a $30$-meter propagation path between the AP and another IoT device (IoT device $2$). The emulation was done by inserting signal delay blocks \cite{gnuradio_signal_delay} in the GNURadio TX paths and RX paths of the IoT devices to delay the incoming and outgoing signals (see Fig. \ref{fig:signal_delay}). Specifically, the signal delay blocks in the GNURadio TX paths emulate the propagation delays ${d_{i,0}}$ and the delay blocks in the GNURadio RX paths emulate ${d_{0,i}}$. No modification was done in the AP's flowgraph. For comparison, we also investigated the performance of the scheme that does not compensate for the propagation delay. 

\begin{figure}
\centering \scalebox{0.4}{\includegraphics{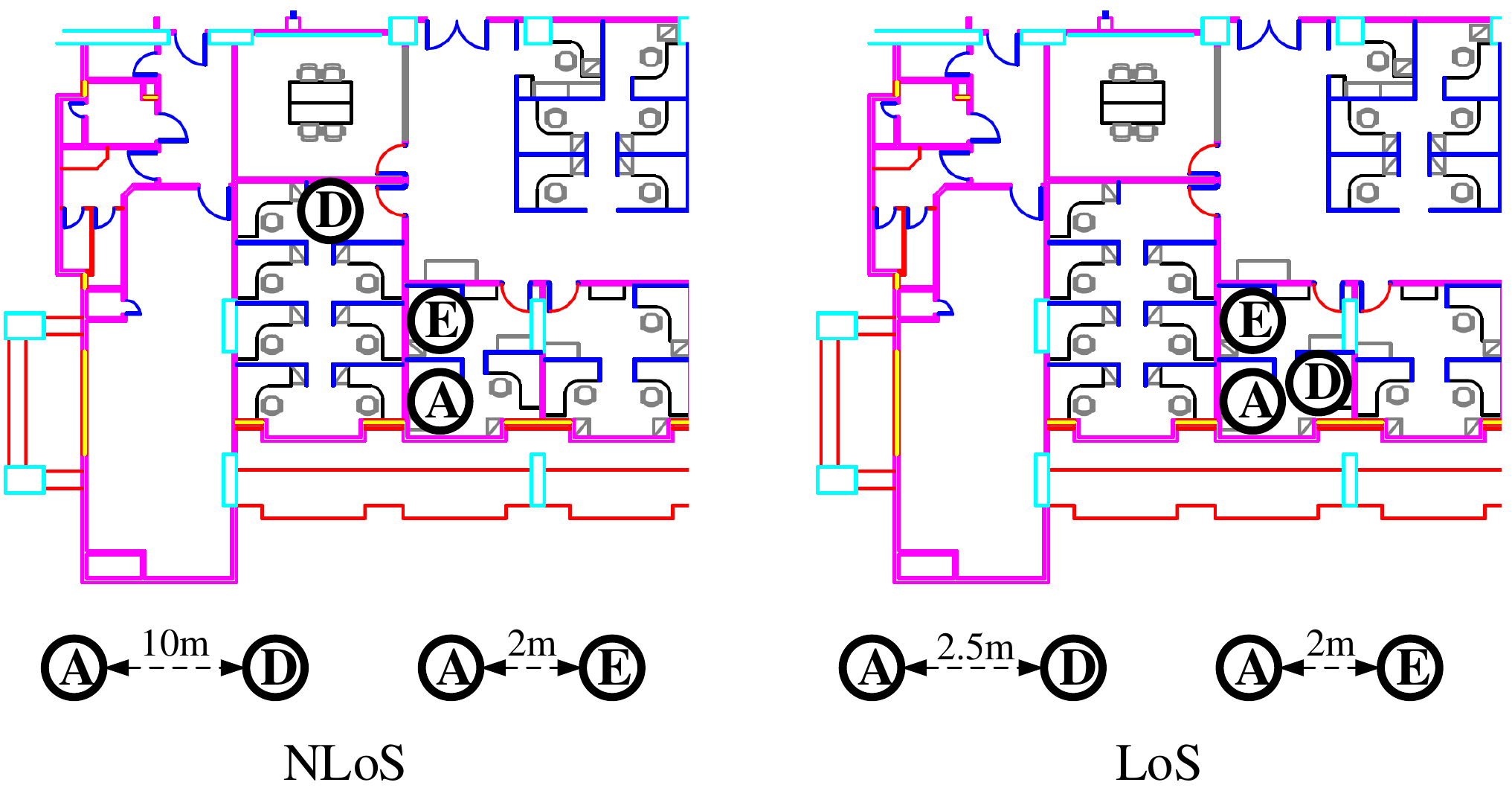}}
\caption{\textcolor{black}{The deployed locations of three sets of ``PC+USRP'' in LoS and NLoS. A is the AP and E, D are the IoT devices. The distances among them are also shown.}}\label{fig:los_nlos}
\end{figure}

Meanwhile. to evaluate the robustness of the synchronization algorithm, the above experiments were carried out in Line-of-Sight (LoS) as well as Non-Line-of-Sight (NLoS) environments. For the Non-Line-of-Sight environment, we put the AP and IoT devices in different rooms where there is a wall blocking the direct paths between them (see Fig. \ref{fig:los_nlos}).

\begin{figure}
\centering \scalebox{0.44}{\includegraphics{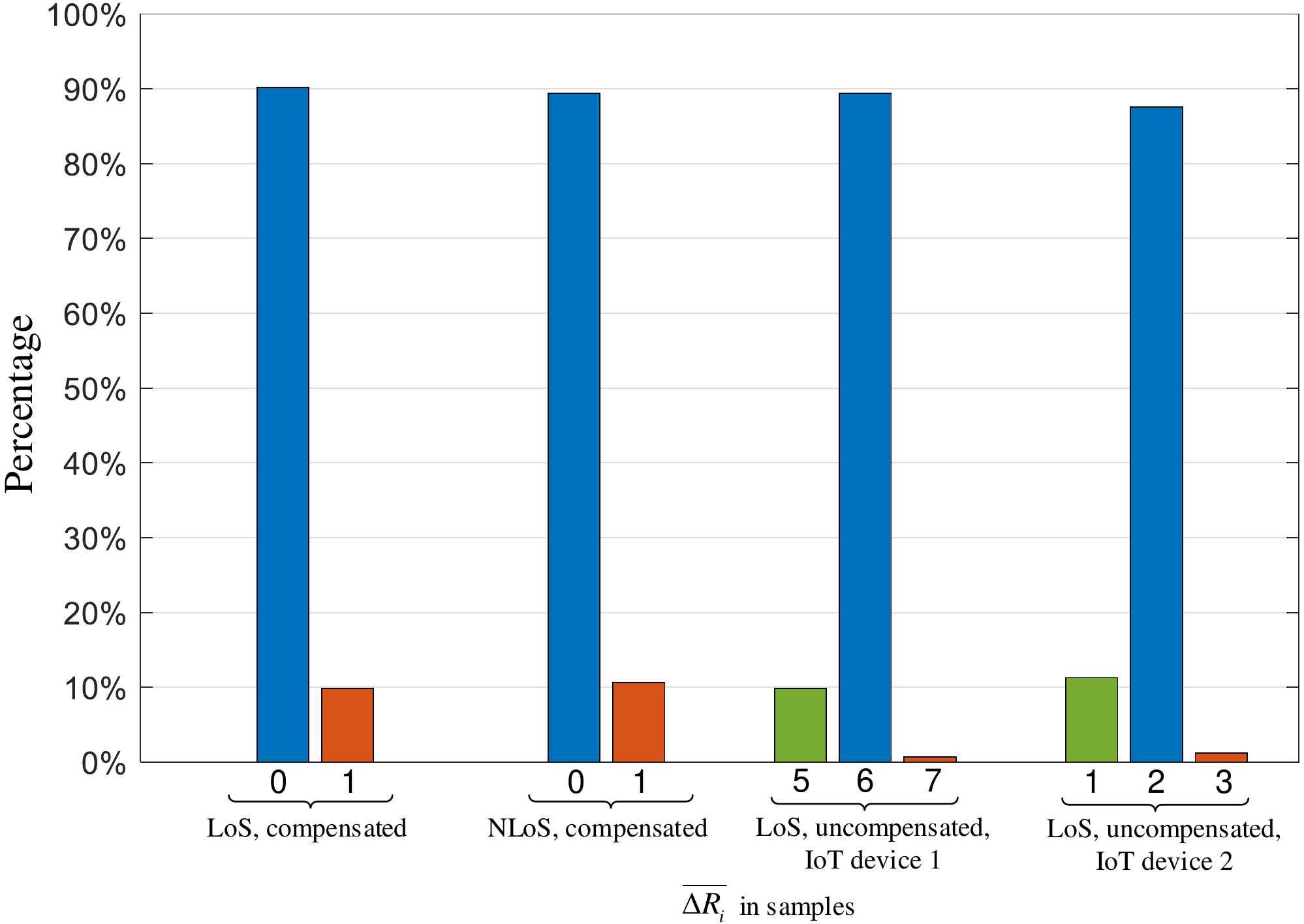}}
\caption{\textcolor{black}{The percentage of achieved $\overline {\Delta {R_i}} $ in the system in three different cases: (1) LoS; (2) NLoS; (3) LoS without propagation delay compensation.}}
\label{fig:sample_drift}
\end{figure}

Ten rounds of tests with ${10^3}$ frames per round were run in each case, and each IoT device occupied one slot in a frame. We then took an average of $\Delta R_{i,j}^k$ over all the frames for each IoT device in each case. We use $\overline {\Delta {R_i}} $ to denote the average value of $\Delta R_{i,j}^k$. \textcolor{black}{The results of propagation delay compensation and robustness are shown in Fig. \ref{fig:sample_drift}. All three cases have the same setting of propagation delays $d_{0,1}=d_{1,0}=3\times {10^{-7}}s$, $d_{0,2}=d_{2,0}=1\times {10^{-7}}s$. The x-axis is the absolute difference between the actual timestamps and the expected timestamps $\overline {\Delta {R_i}} $, and the y-axis is the percentage of that value for $\overline {\Delta {R_i}} $.} Note that the resolution of our misalignment measurement is $1$ sample ($0.1\mu s$). Thus, a measured misalignment of $0$ corresponds to misalignment of $-0.5$ to $+0.5$ samples and a measured misalignment of $1$ corresponds to misalignment of $-1.5$ to $-0.5$ or $0.5$ to $1.5$ samples. Fig. \ref{fig:sample_drift} shows that our system can align the slot boundaries to within $ \pm 0.5$ samples $90\%$ of the time and to within $ \pm 1.5$ samples $100\%$ of the time for both LoS and NLoS environments. If the emulated propagation delay is not compensated, on the other hand, additional misalignment corresponding to the uncompensated round-trip delay will be introduced.

\textcolor{black}{We now evaluated the performance of our synchronization algorithm when the IoT devices adopt low-cost oscillators. To that end, we consider a typical low-cost oscillator used for IoT devices \cite{iot_clock_website}. According to the data sheet \cite{iot_clock_website}, the oscillator provides $\pm 50ppm$ frequency stability, which is $20$ times less accurate than the X310's high-end Temperature Compensate Crystal Oscillator (TCXO) that has a frequency stability of $\pm 2.5ppm$.}

\textcolor{black}{Because the low-end oscillator cannot be directly used as the clock source for the USRP, we emulated the effect of the low-cost oscillator by performing baseband digital signal processing at the TX path. In OFDM transmission, carrier frequency offset (CFO) is the most severe effect caused by the frequency instability. Because the low-end oscillator is 20 times less accurate than the X310's oscillator, the CFO between the transmitter and the receiver can increase by up to $20$ times. Given that our system operates at $f = 2.418GHz$, the frequency variation is $120.9kHz$. The maximum CFO, therefore, equals to $120.9kHz \times 2 = 241.8kHz$. We thus artificially insert frequency offsets to the baseband samples $x[n]$ by ${e^{j2\pi \Delta fn}}$ (i.e., the baseband samples passed to the RF board are $\tilde x[n] = x[n] \cdot {e^{j2\pi \Delta f n}}$, where $\Delta f$ is the double-sided phase deviation per sample and $\Delta f = 241.8kHz \times {\textstyle{1 \over {10MHz}}} = 0.02418$).}

\begin{figure}
	\centering \scalebox{0.53}{\includegraphics{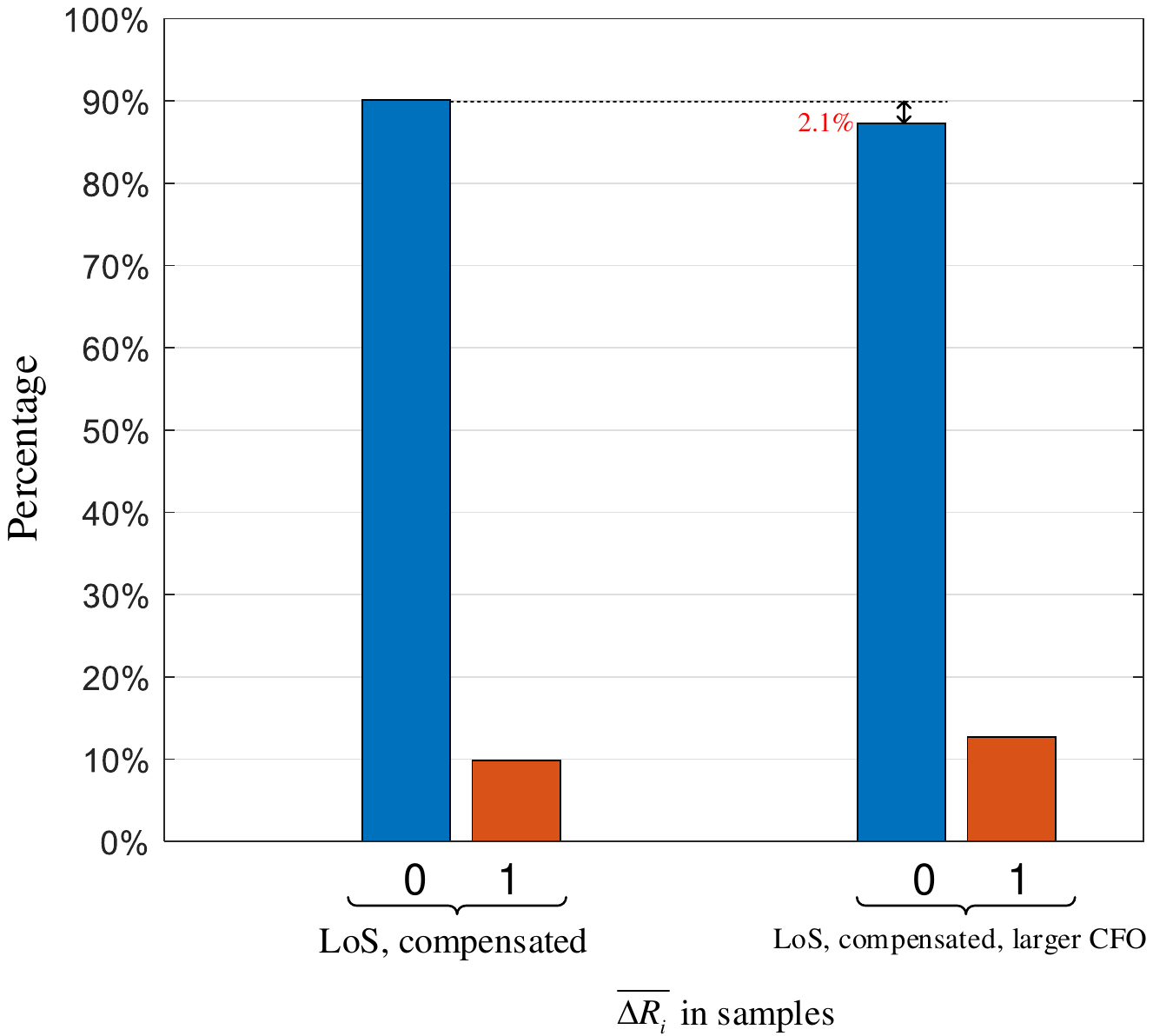}}
	\caption{\textcolor{black}{Comparison of the synchronization mechanism's performance with high-end oscillators and low-cost oscillators.}}
	\label{fig:low_cost_perf}
\end{figure}

\textcolor{black}{The experiments were carried out using the same parameters as in Table \ref{table:exp}. Ten rounds of tests with ${10^3}$ frames per round were run in each case. The result is shown in Fig. \ref{fig:low_cost_perf}, in which the performance of RTTS-SDR in the LoS scenario with propagation delay compensation is used as the benchmark. Thanks to the robustness of our packet detection algorithm and the CFO correction in our system, the extra CFO introduced by the low-cost oscillator does not lead to substantial synchronization performance degradation. Specifically, the percentage of perfect synchronization (i.e., slot boundaries are aligned within $\pm 0.5$ samples) only decreases by $2.1\% $, and $\overline {\Delta {R_i}} \le 1$ for $100\% $ of the time. Therefore, we can conclude that the low-cost oscillator will not substantially affect the synchronization performance of the system.}

\subsection{Latency of packet delivery}	\label{subsec:latency_pkt_exp}
To demonstrate the benefits of our {\textit{Just-in-time}} algorithm, we performed tests to measure end-to-end latency. Specifically, we measured the round-trip time of an IoT device delivering a packet to the AP followed by the AP delivering a packet back to the IoT device. Just before preparing a packet for transmission, the IoT device marks down its PC time as the packet's transmission PC time ${t_{{\rm{P - TX}}}}$. After the AP's PC receives and decodes the packets, it prepares a feedback packet and sends it back to the IoT device. When the IoT device receives the feedback packet from the AP, it checks the transmission PC Time $t_{{\rm{P - RX}}}$ of the packet and compute the round-trip time by ${t_{{\rm{RTT}}}} = {t_{{\rm{P - RX}}}} - {t_{{\rm{P - TX}}}}$.

Before running the tests, we measured the RTT between PC and the USRP of the IoT device $i$, i.e., ${\delta _{{\rm{RTT}},i}}$, by running the ``PING'' test. Because the raw samples were transmitted between PC and the USRP with bursty transmission, we let the PING test ran in a bursty way. The PING test was done for $10$ rounds, and in each round we sent $1000$ packets with the transmission interval equals to $1ms$. The PING test shows that the mean RTT, ${\delta _{{\rm{RTT}},i}}$, is $1.154ms$ and the deviation, ${\sigma _{{\rm{RTT}},i}}$, is $0.812ms$. Therefore, the time we should send in advance based on (\ref{eqa:new_packet_tx_time}) can be set to be $2ms$, assuming $\beta  = 1$ and a small amount of added time for packet preparation.

Again, ${T_{{\rm{adv}}}}$ is the time for the PC to ``wake ahead''. For our {\textit{Just-in-time}} algorithm, we set ${T_{{\rm{adv}}}} = 2ms$. For benchmarking, we also ran the tests with ${T_{{\rm{adv}}}} = 10ms$ which is approximately half the duration of a frame, and ${T_{{\rm{adv}}}} = 20ms$, which is approximately the duration of a frame. Each test consists of $10^6$ pairs of packets between the IoT device and the AP. Based on the statistics of these packet, we obtain the $99$-percentile and $99.99$-percentile RTT.

\begin{figure}
\centering \scalebox{0.62}{\includegraphics{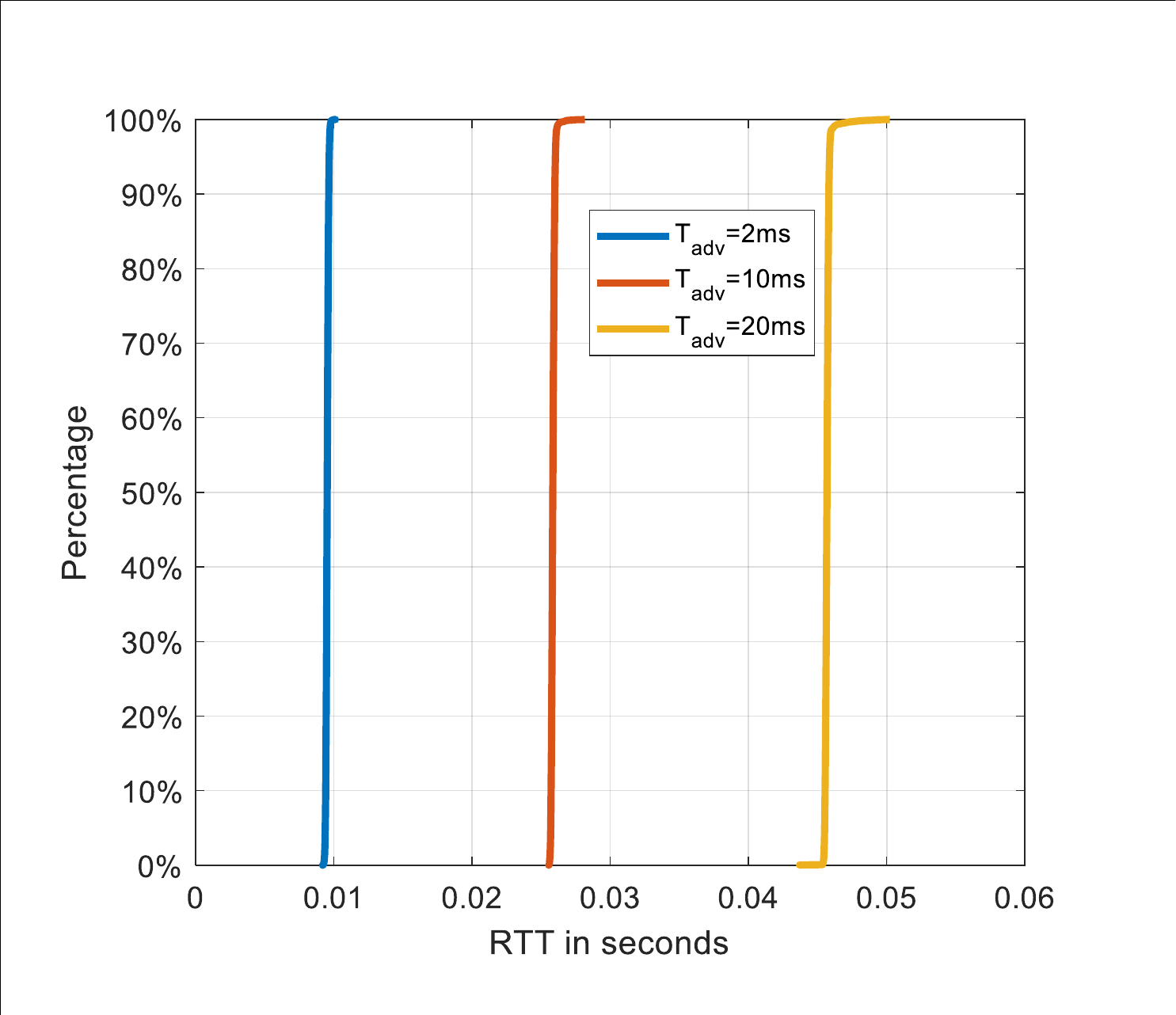}}
\caption{The RTT of delivering a packet with different ${T_{{\rm{adv}}}}$.}\label{fig:rtt}
\end{figure}

As shown in Fig. \ref{fig:rtt}, ${T_{{\rm{RTT}}}}$ depends significantly on ${T_{{\rm{adv}}}}$. For ${T_{{\rm{adv}}}} = 2ms$, the $99$-percentile RTT is $9.97ms$ and the $99.99$-percentile is $10.39ms$. For ${T_{{\rm{adv}}}}=10ms$, the $99$-percentile RTT is $26.16ms$ and the $99.99$-percentile RTT is $27.49ms$. And for ${T_{{\rm{adv}}}}=20ms$, the $99$-percentile RTT is $46.19ms$ and the $99.99$-percentile is $49.28ms$.

\begin{figure}
\centering \scalebox{0.58}{\includegraphics{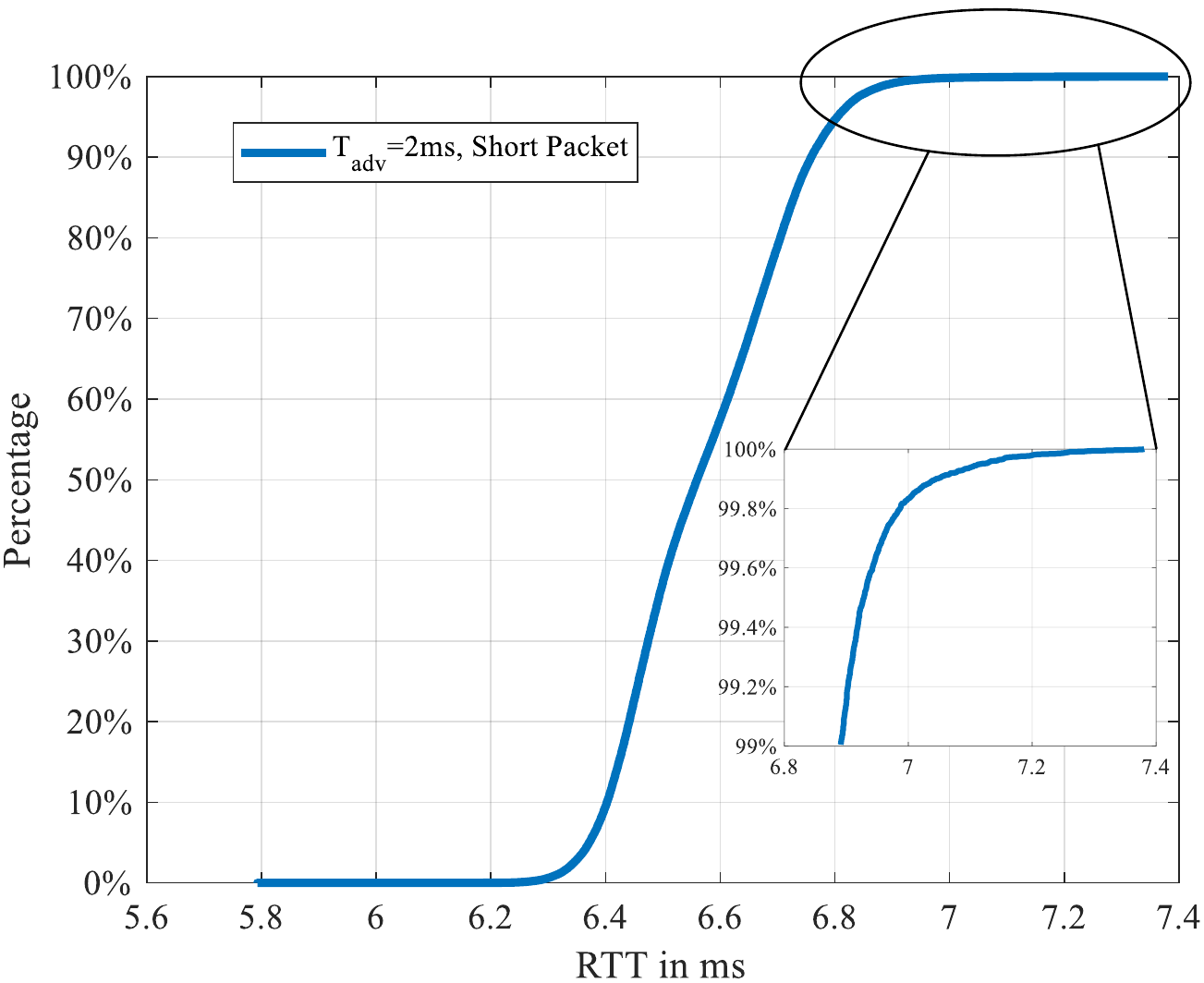}}
\caption{\textcolor{black}{The RTT of delivering a short packet with ${T_{{\rm{adv}}}}=2ms$.}}\label{fig:rtt_short}
\end{figure}

We further explored the ${T_{{\rm{RTT}}}}$ of short packets. IoT applications with low-latency requirements typically need to transmit very little data. Thanks to the reconfigurability of RTTS-SDR (see Appendix \ref{appendix:reconfigurability}), we could easily reduce the length of payload in a packet from $128$ to $12$. The PHY-layer uses BPSK modulation and $1/2$ convolutional code, giving a packet length of $36$ bytes, which is close to the packet size of $32$ bytes defined in 5G for ultra-reliable low-latency communication (URLLC) \cite{3gpp.38.913}. The guard time between two consecutive slots is also reduced from $360$ samples to $80$ samples. We set ${T_{{\rm{adv}}}}$ to $2ms$. To explore the $99.9999$-percentile RTT, we ran the tests with $10^7$ pairs of short packets between the IoT device and the AP. As shown in Fig. \ref{fig:rtt_short}, the $99$-percentile RTT is $6.90ms$, the $99.99$-percentile RTT is $7.24ms$, and the $99.9999$-percentile RTT is $7.37ms$. Indeed, the RTTs of all packets are bounded by $7.5ms$. This implies that the one-way latency is bounded by $3.75ms$.

In the context of a control system in which the controller is connected to the AP and the sensor and the actuator are connected to the IoT device, the above round-trip delay corresponds to the feedback-loop latency of the control system. Our system can guarantee a feedback loop latency bounded by $7.5ms$.


%
{
\section{Conclusions}
This paper puts forth a real-time time-slotted system on the USRP-SDR platform for time-sensitive wireless networks. Specifically, we proposed two new techniques to handle the issues raised by the latency between the PC and the USRP: (i) sample counting in the receive path for time-slot synchronization; (ii) {\textit{Just-in-time}} algorithm to time the forwarding of a packet from the PC to the USRP in the transmit path to achieve low latency. With these techniques, the system can achieve sub-microsecond time synchronization ($100ns$) among nodes and $3.75$-ms end-to-end latency. Overall, our system can fulfill part of the URLLC defined in 5G (e.g., medium-voltage electric power distribution grid, augmented reality, and mobile panel control panels with safety functions \cite[pp.159]{3gpp.22.804}). It demonstrates the viability of building the time-sensitive wireless IoT networks on the SDR platform. 



\appendix
\section*{Reconfigurability\footnotemark}	\label{appendix:reconfigurability}
\footnotetext{Readers who are interested in the software code of RTTS-SDR can send a request to \cite{source_code_web}. We have attempted to make RTTS-SDR reconfigurable for other systems than the TDMA system described in this paper, and we are interested in feedback from users who would like to try out RTTS-SDR.}

Many of the parameters in the PHY-layer of RTTS-SDR are reconfigurable. Specifically, the following parameters can be changed easily in our system: (1) Subcarrier mapping in the OFDM modulation and demodulation; (2) Length of the CP; and (3) Packet length; and (4) Bitrates. In this appendix, we briefly introduce how the system provides support to each of the reconfigurable parameters.

\textit{Subcarrier mapping}. A changeable subcarrier mapping allows the system to fit into wireless channels of different spectrum characteristics. For example, when the unused spectrum is discontinuous, the system can disable some of its subcarriers so that it will not interfere with other existing wireless systems that already occupy the used spectrum of those subcarriers. Packet detection algorithm, which has been modified to be generic, is capable of detection any kind of subcarrier mapping.

\textit{Length of OFDM CP}. The length of CP is an essential parameter for the scenarios where the delay spread is a concern. For our time-slotted system, the length of CP ${N_{{\rm{CP}}}}$ can vary from 0 to the length of FFT/IFFT ${N_{{\rm{FFT}}}}$. Note that the CP may be a large overhead if a system is set with a large ${N_{{\rm{CP}}}}$ but runs in a small-delay-spread environment. The required CP length depends on the coverage of the system to be prototyped.

\textit{Packet length}. The length of packets in a system determines the scope of applications of that system. Our system supports packet lengths starting from 0 (no payload) to any positive value.

\textit{Bitrates}. Our system supports all the bitrates in 802.11a/g/n and it is also capable of changing the bitrate packet-by-packet. Users can easily change the bitrate of a packet by giving different parameters to the API.

\ifCLASSOPTIONcaptionsoff
  \newpage
\fi

\bibliographystyle{IEEEtran}
\bibliography{References}

\end{document}